\documentclass[a4paper,11pt]{article}
\usepackage{jheppub}
\usepackage[utf8]{inputenc}
\usepackage{graphicx,cancel,float}
\usepackage{amssymb}
\usepackage{textcomp}
\usepackage{amsmath}
\usepackage{tabularx}
\usepackage{bm}
\usepackage{times}
\usepackage{color}
\usepackage{slashed}
\usepackage{multirow}
\usepackage{verbatim}
\usepackage{epsfig,epstopdf}

\def\lsim{\mathrel{\raise.3ex\hbox{$<$\kern-.75em\lower1ex\hbox{$\sim$}}}}
\def\gsim{\mathrel{\raise.3ex\hbox{$>$\kern-.75em\lower1ex\hbox{$\sim$}}}}

\newcommand{\calO}{{\mathcal{O}}}

\newcommand{\T}{{\rm T}}
\newcommand{\GeV}{{\rm GeV}}

\usepackage[table]{xcolor}
\usepackage{colortbl}
\definecolor{mygray}{gray}{.95}

\preprint{}
\subheader{\it \today}

\title{General neutrino interactions with sterile neutrinos in light of coherent neutrino-nucleus scattering and meson invisible decays}

\author[a]{Tong Li}
\emailAdd{litong@nankai.edu.cn}
\affiliation[a]{School of Physics, Nankai University, Tianjin 300071, China}
\author[b]{Xiao-Dong Ma}
\emailAdd{maxid@phys.ntu.edu.tw}
\affiliation[b]{Department of Physics, National Taiwan University, Taipei 10617, Taiwan}
\author[c]{Michael A. Schmidt}
\emailAdd{m.schmidt@unsw.edu.au}
\affiliation[c]{School of Physics, The University of New South Wales, Sydney, New South Wales 2052, Australia}

\abstract
{In this work we study the current bounds from the CE$\nu$NS
process and meson invisible decays on generic neutrino interactions with sterile
neutrinos in effective field theories. The interactions between quarks and
left-handed SM neutrinos and/or right-handed neutrinos are first described by
the low-energy effective field theory (LNEFT) between the electroweak scale and
the chiral symmetry breaking scale. We complete the independent operator basis
for the LNEFT up to dimension-6 by including both the lepton-number-conserving
(LNC) and lepton-number-violating (LNV) operators involving right-handed
neutrinos. We translate the bounds on the LNEFT Wilson coefficients from the
COHERENT observation and calculate the branching fractions of light meson
invisible decays. The bounds on LNEFT are then mapped onto the SM effective field theory with sterile
neutrinos (SMNEFT) to constrain new physics above the electroweak scale.
We find that the meson invisible decays can provide the only sensitive probe
for $\tau$ neutrino flavor component and $s$ quark component in the
quark-neutrino interactions involving two (one) active neutrinos and for the effective operators without
any active neutrino fields. The CE$\nu$NS process places the most stringent
bound on all other Wilson coefficients. By assuming one dominant Wilson
coefficient at a time in SMNEFT and negligible sterile neutrino mass, the most
stringent limits on the new physics scale are $2.7-10$ TeV from corresponding
dipole operator in LNEFT and $0.5-1.5$ TeV from neutrino-quark operator in
LNEFT.
}

\arxivnumber{arXiv:2005.01543}

\begin{document}

\maketitle
\setcounter{page}{2}

\newpage

\section{Introduction}
\label{sec:Intro}

The coherent elastic neutrino-nucleus scattering (CE$\nu$NS) process has been
observed by the COHERENT experiment at the $6.7\sigma$
level~\cite{Akimov:2017ade}. The Standard Model (SM) predicts the CE$\nu$NS
process through the $Z$ boson exchange~\cite{Freedman:1973yd} and the
observation is consistent with the SM at the $1\sigma$ level.  In the COHERENT
experiment, the spallation neutron source produces prompt $\nu_\mu$ and delayed
$\bar{\nu}_\mu, \nu_e$ which reach the low-background CsI detector. Besides the
active neutrinos through the weak neutral current in the SM, any neutrino
flavors including light right-handed (RH) neutrinos\footnote{RH neutrinos refer to sterile neutrinos which do not carry any SM gauge charges.} can be produced in the final state of
the CE$\nu$NS process. The COHERENT observation thus provides us an opportunity to
explore the new physics (NP) associated with generic neutrino interactions in the presence of
light RH neutrinos.

CE$\nu$NS occurs when the transferred momentum during the neutrino scattering
off a nucleus is smaller than the inverse of the nuclear radius. Thus, the
relevant neutral currents can be well described by an effective field theory
(EFT) below the electroweak (EW) scale.
The low-energy effective field theory (LEFT) is an EFT defined below the
electroweak scale $\Lambda_{\rm EW}\sim 10^2$ GeV.
In the LEFT, the dynamical degrees
of freedom are the SM charged and neutral leptons and light quarks excluding
the heavy
top quark. They respect the unbroken gauge symmetries $SU(3)_c\times
U(1)_{\rm em}$ after integrating out the Higgs boson $h$, weak gauge bosons $W, Z$ and the top quark $t$ in the SM. The basis of LEFT operators up to dimension-6 (dim-6) has
been written down in Ref.~\cite{Jenkins:2017jig}.
If the LEFT is extended by right-handed (RH) neutrinos $N$, the corresponding effective field theory
is named as LNEFT. An independent subset of lepton-number-conserving (LNC)
operators with RH neutrinos $N$ at dim-6 in LNEFT was given in
Ref.~\cite{Chala:2020vqp}. In this paper we construct the additional
 lepton-number-violating (LNV) operators up to dim-6
involving $N$ which may or may not break the baryon
  number. Together with those in Refs.~\cite{Jenkins:2017jig,Chala:2020vqp},
  they make up the complete and independent operator basis for the LNEFT up
  to dim-6\footnote{The basis of LNV operators was also constructed in the journal version of Ref.~\cite{Chala:2020vqp}}. Moreover, to connect to NP above the
  electroweak scale, we match the LNEFT to the SM
    effective field theory extended by RH neutrinos $N$ (SMNEFT)
    at the electroweak scale~\cite{
Grzadkowski:2010es,Lehman:2014jma, Liao:2016hru,Liao:2016qyd}. The SMNEFT
respects the SM gauge group $SU(3)_c\times SU(2)_L\times U(1)_Y$ and
describes the physics above the electroweak scale up to the NP scale. In this
paper we revise the SMNEFT operator basis involving $N$ in
Ref.~\cite{Liao:2016qyd} by changing some notations.

Recently, Refs.~\cite{Lindner:2016wff,AristizabalSierra:2018eqm,Altmannshofer:2018xyo,Bischer:2019ttk,Chang:2020jwl,Han:2020pff} considered constraints from the CE$\nu$NS process
on the LNC operators in LNEFT and SMNEFT. With the complete basis of LNEFT
and SMNEFT, however, we can perform a comprehensive study of the constraints
on both the LNC and LNV cases and investigate the implication for NP above
the electroweak scale.

Besides the CE$\nu$NS process, invisible decays of light mesons can give
additional information for the nature of neutrinos. In the SM, the decay
rates of $\pi^0, \eta, \eta'\to \nu\bar{\nu}$ are helicity suppressed and
those of light vector mesons ($\phi$ and $\omega$) are also extremely small. Thus, the observation
of any of these meson invisible decays would clearly indicate the existence
of NP~\cite{Gninenko:2014sxa,Gninenko:2015mea,Abada:2016plb,Gao:2018seg}.
Moreover, they can provide the only sensitive probe for some flavor
components in the quark-neutrino interactions~\cite{Li:2019fhz} and for the
effective operators without any active neutrino fields. In this work we study
the correlation and complementarity of the CE$\nu$NS process and meson invisible
decay for the bound on generic neutrino interactions with RH neutrinos in
the frameworks of LNEFT and SMNEFT.

The paper is outlined as follows. In Sec.~\ref{sec:GNI}, we describe the
generic neutrino-photon/quark operators in the LNEFT basis. The LNEFT operators are
then matched to the SMNEFT. We derive the general constraints on LNEFT Wilson
coefficients (WCs) in both LNC and LNV cases from the CE$\nu$NS process in
Sec.~\ref{sec:CEnuNS}. In Sec.~\ref{sec:inv}, we give the analytical
expressions for the invisible decay branching fractions of light mesons.
Sec.~\ref{sec:Num} shows our numerical results and the lower bounds on the NP
scales in the SMNEFT. Our conclusions are drawn in Sec.~\ref{sec:Con}. Some
calculational details are collected in the Appendixes.

\section{General neutrino interactions with RH neutrinos}
\label{sec:GNI}

The main focus of this work is on low-energy processes CE$\nu$NS and light meson invisible decays. Thus, we will start from the framework of LNEFT. The LNEFT is defined below the electroweak scale $\Lambda_\textrm{EW}$ and its dynamical degrees of freedom include the SM light particles excluding $h,W,Z,t$ and an arbitrary number of RH neutrinos $N$. The power counting of LNEFT is determined by both the NP scale $\Lambda_{\rm NP}$ and the electroweak scale $\Lambda_{\rm EW}$. The LNEFT consists of dim-3 fermion mass terms, dim-4 kinetic terms and higher dimensional operators $\calO_{i,L}^{(d)}(d\geq 5)$ (dim-$d$) built out of those light fields and satisfies the $SU(3)_c\times U(1)_{\rm em}$ gauge symmetry. The LNEFT Lagrangian is
\begin{align}
\mathcal{L}_{\rm LNEFT}=\mathcal{L}_{\rm d\leq 4}+ \sum_i \sum_{d\geq 5} C_{i,L}^{(d)} \calO_{i,L}^{(d)}\;,
\end{align}
where $C_{i,L}^{(d)}$ is the Wilson coefficient of operator $\calO_{i,L}^{(d)}$. Generally, the Wilson coefficients $C_{i,L}^{(d)}$ scale as $\Lambda_{\rm EW}^{n+4-d}/\Lambda_{\rm NP}^{n}$ with integer $n\geq 0$. In Appendix~\ref{sec:LNEFTbasis} we construct the complete and independent operator basis involving RH neutrinos $N$ up to dim-6 in the LNEFT for the study of generic neutrino interactions. In this work we consider the non-renormalisable dim-5 magnetic dipole operators and dim-6 four-fermion operators.

We assume the LNEFT is a low-energy version of the SMNEFT which is defined above the electroweak scale.
In the SMNEFT, the renormalizable SM Lagrangian is extended by the RH neutrino sector and a tower of higher dimensional effective operators $\calO_i^{(d)}$ with increasing canonical dimension $d\geq 5$.
The importance of these operators is measured by the Wilson coefficients $C_i^{(d)}$ with decreasing relevance
\begin{eqnarray}
\mathcal{L}_{\rm SMNEFT}= \mathcal{L}_{\rm SM+N} + \sum_i \sum_{d\geq 5} C_i^{(d)} \calO_i^{(d)} \; ,
\end{eqnarray}
where $\mathcal{L}_{\rm SM+N}$ is the renormalizable SM Lagrangian extended by RH neutrinos $N$.
Generally, each Wilson coefficient $C_i^{(d)}$ is associated with a NP scale $\Lambda_{\rm NP}=(C_i^{(d)})^{1/(4-d)}$. For a given NP model, it can be precisely expressed as the function of the parameters in the NP model through matching and renormalization group running procedures. In Appendix~\ref{sec:SMNEFTbasis} we collect the relevant SMNEFT operators used in our analysis for the generic neutrino interactions.

\subsection{General neutrino operators in LNEFT basis}
\label{sec:LEFT}
The generic neutrino operators entering the framework of LNEFT respect SU$(3)_{\rm
c}\times$U$(1)_{\rm em}$ gauge symmetry and are constructed by a neutrino
bilinear coupled to the photon field strength tensor or SM quark bilinear currents.\footnote{The operators involving charged leptons are not related to the processes of interest and thus we do not consider them here.}
In the basis of LNEFT for neutrinos, the dim-5 neutrino-photon and dim-6 neutrino-quark operators with lepton number conservation (LNC, $|\Delta L|=0$) are given by~\cite{Jenkins:2017jig}
\begin{align}
\calO_{\nu N F}&=(\overline{\nu}\sigma_{\mu\nu}N) F^{\mu\nu} +h.c.\; ,
\\
\calO_{q\nu 1}^{V}&=(\overline{q_L}\gamma_\mu q_L)(\overline{\nu}\gamma^\mu\nu)\; , &
\calO_{q\nu 2}^{V}&=(\overline{q_R}\gamma_\mu q_R)(\overline{\nu}\gamma^\mu\nu)\; ,
\label{LNCqnu}
\\
\calO_{qN 1}^{V}&=(\overline{q_L}\gamma_\mu q_L)(\overline{N}\gamma^\mu N)\; , &
\calO_{qN 2}^{V}&=(\overline{q_R}\gamma_\mu q_R)(\overline{N}\gamma^\mu N)\; ,
\label{LNCqN}
\\
\calO_{q\nu N1}^S&= (\overline{q_L} q_R)(\overline{\nu} N)+h.c.\; ,&
\calO_{q\nu N2}^S&= (\overline{q_R} q_L)(\overline{\nu} N)+h.c.\; ,
\label{LNCqnuN}
\\
\calO_{q\nu N}^T&= (\overline{q_L} \sigma^{\mu\nu}q_R)(\overline{\nu}\sigma_{\mu\nu} N)+h.c.\; ,
\label{LNCqnuNT}
\end{align}
where $F_{\mu\nu}$ is the electromagnetic field strength tensor, $q$ can be either up-type quarks $u_i$ or down-type quarks $d_i$, $\nu_i$ are active left-handed neutrinos, and $N_i$ are RH neutrinos. The quark fields and the RH neutrino fields are in the mass basis, while the left-handed (LH) neutrino fields are in the flavor basis.
Both $\nu_i$ and $N_i$ carry lepton number $L(\nu_i)=L(N_i)=+1$.
The flavors of the two quarks and those of the two neutrinos in the above operators can be different although we do not specify their flavor indexes here.
For the notation of the Wilson coefficients, we use the same subscripts as the operators, for instance $C_{q\nu1}^{V,pr\alpha\beta}$ together with $\calO_{q\nu1}^{V,pr\alpha\beta}$, where $p,r$ denote the quark flavors and $\alpha,\beta$ are the neutrino flavors.
We demand the vector operators to be hermitian, i.e. $C_X^{V, pr\alpha\beta}=C_X^{V, rp\beta\alpha\ast}$ with
$X=q\nu 1, q\nu 2, qN1, qN2$, to ignore the $h.c.$ in Eqs.~(\ref{LNCqnu}) and (\ref{LNCqN}).

The relevant dim-5 and dim-6 operators which induce lepton number violation (LNV, $|\Delta L|=2$) are
\begin{align}
\calO_{\nu \nu F}&=(\overline{\nu^C}\sigma_{\mu\nu}\nu) F^{\mu\nu} +h.c.\; , &
\calO_{NN F}&=(\overline{N^C}\sigma_{\mu\nu}N) F^{\mu\nu} +h.c.\; ,
\\
\calO_{q\nu N1}^V&= (\overline{q_L}\gamma_\mu q_L)(\overline{\nu^C}\gamma^\mu N)+h.c.\; , &
\calO_{q\nu N2}^V&=(\overline{q_R}\gamma_\mu q_R)(\overline{\nu^C}\gamma^\mu N)+h.c.\; ,
\label{LNVqnuN}
\\
\calO_{q\nu 1}^S&= (\overline{q_R} q_L)(\overline{\nu^C} \nu)+h.c.\; ,&
\calO_{q\nu 2}^S&= (\overline{q_L} q_R)(\overline{\nu^C} \nu)+h.c.\; ,
\label{LNVqnu}
\\
\calO_{qN 1}^S&= (\overline{q_R} q_L)(\overline{N^C} N)+h.c.\; ,&
\calO_{qN 2}^S&= (\overline{q_L} q_R)(\overline{N^C} N)+h.c.\; ,
\label{LNVqN}
\\
\calO_{q\nu}^T&= (\overline{q_R} \sigma^{\mu\nu}q_L)(\overline{\nu^C}\sigma_{\mu\nu} \nu)+h.c.\;, &
\calO_{qN}^T&= (\overline{q_L} \sigma^{\mu\nu}q_R)(\overline{N^C}\sigma_{\mu\nu} N)+h.c.\; .
\label{LNVT}
\end{align}
Note that the Wilson coefficients of the scalar operators are symmetric in the neutrino indices and the dipole and tensor operators are antisymmetric in the neutrino indices. Thus in particular the operators with the tensor neutrino current $\overline{\nu_\alpha^C} \sigma^{\mu\nu} \nu_\beta$ or $\overline{N_\alpha^C} \sigma^{\mu\nu} N_\beta$ vanish for identical neutrino flavors ($\alpha=\beta$).

LNEFT is a valid description between the electroweak scale $\Lambda_{\text{EW}}=m_W$ and chiral symmetry breaking scale $\Lambda_\chi\simeq 1~\GeV$ for the low-energy neutrino-quark interactions. There are large logarithms produced by the ratio of the two scales in the perturbative expansion which can be resummed by solving the relevant renormalization group equations. In our case, the leading order contribution comes from the one-loop QCD and QED corrections. The vector (and axial-vector) current operators are not renormalized at one-loop level because of the QED and QCD Ward identities. However, the scalar and tensor current operators are renormalized and their one-loop renormalization group equations for the corresponding Wilson coefficients are~\cite{Jenkins:2017dyc}
\begin{align}
\mu{d \over d\mu}C^S_q&=-\left( {\alpha_s \over 2\pi}3C_F+ {\alpha \over 2\pi}3Q^2_q\right)C^S_q, &
&C^S_q\in \{C_{q\nu\,N1}^{S}, C_{q\nu\,N2}^{S},C_{q\nu1}^{S},C_{q\nu2}^{S},C_{qN1}^{S},C_{qN1}^{S}\}\;, \nonumber
\\
\mu{d \over d\mu}C^T_q&=\left( {\alpha_s \over 2\pi}C_F+ {\alpha \over 2\pi}Q^2_q\right)C^T_q, &
&C^T_q\in \{C_{q\nu N}^{T},C_{q\nu}^{T}, C_{qN}^{T} \}\;,
\end{align}
where $C_F=(N^2_c-1)/2N_c=4/3$ with $N_c=3$ is the second Casimir invariant of the color group $SU(3)_c$, $Q_q$ is the electric charge of quark field $q$ in the corresponding operator in unit of positron's charge $e$, and $\alpha=e^2/(4\pi) (\alpha_s=g_s^2/(4\pi))$ is the (strong) fine structure constant. The solutions for the above equations are straightforward and are given by
\begin{align}
C^S_q(\mu_1)&=\left({ \alpha_s(\mu_2)\over \alpha_s(\mu_1)} \right)^{3C_F/b}\left({ \alpha(\mu_2)\over \alpha(\mu_1)} \right)^{3Q^2_q/ b_e}C^S_q(\mu_2)\;, \nonumber
\\
C^T_q(\mu_1)&=\left({ \alpha_s(\mu_2)\over \alpha_s(\mu_1)} \right)^{-C_F/ b}\left({ \alpha(\mu_2)\over \alpha(\mu_1)} \right)^{-Q^2_q/b_e}C^T_q(\mu_2)\;,
\end{align}
between two scales $\mu_1$ and $\mu_2$.
Here $b=-11+2/3n_f$ with $n_f$ being the number of active quark flavors between scales $\mu_1$ and $\mu_2$, and  $b_e = \sum_i \tfrac43 (N_c)_i Q_i^2=4(3 n_\ell + 4 n_u +  n_d)/9$ with $n_{\ell,u,d}$ being the active number of leptons/up-type quarks/down-type quarks between the two scales. If we take the scale $\mu_1$ ($\mu_2$) to be $\Lambda_\chi$ ($\Lambda_{\text{EW}}$), after including quark and lepton threshold effects, the numerical results are
\begin{align}
C^S_u(\Lambda_\chi)&=1.67\, C^S_u(\Lambda_\text{EW})\;,&
C^S_d(\Lambda_\chi)&=1.66\, C^S_d(\Lambda_\text{EW})\;, \nonumber
\\
C^T_u(\Lambda_\chi)&=0.85\, C^T_u(\Lambda_\text{EW})\;,&
C^T_d(\Lambda_\chi)&=0.84\, C^T_d(\Lambda_\text{EW})\;.
\end{align}
Compared to the pure QCD running effect in~\cite{Liao:2020roy}, we find the QED correction is almost negligible.
We can see the scalar-type operators are enhanced while the tensor-type operators are suppressed when evolving from the high scale $\Lambda_\text{EW}$ down to the low scale $\Lambda_\chi$.

For the neutrino dipole operators, the renormalization group evolution is
\begin{align}
  \mu\frac{d}{d\mu}C_{iF}^{\alpha\beta}&={ \alpha\over 2\pi} {b_e\over 2}C_{iF}
  - \frac{eN_c}{2\pi^2} \sum_r Q_{q_r} m_{q_r} C_{qi}^{T,rr\alpha\beta} \theta(\mu- m_{q_r})\;,
\label{RGEdipole}
\end{align}
where $C_{iF}\in \{C_{\nu NF}, C_{\nu\nu F}, C_{NNF}\}$ and $C_{qi}^T\in \{C_{q\nu N}^T, C_{q\nu}^T, C_{qN}^T\}$ and $\theta$ is the Heaviside theta function.
It includes the one-loop QED running of $C_{iF}$~\cite{Jenkins:2017dyc} which is the first term on the right-hand side of Eq.~(\ref{RGEdipole}).
Note that, at one-loop order, the renormalization group evolution of the Wilson coefficients of the tensor operators $\calO_{q\nu N}^T$, $\calO_{q\nu}^T$ and $\calO_{qN}^T$ induces a mixing into the dim-5 dipole operators $\calO_{\nu NF}$, $\calO_{\nu\nu F}$ and $\calO_{NNF}$, that is the last term of Eq.~(\ref{RGEdipole})\footnote{The dipole running also receives a similar contribution from the LNV tensor neutrino-charged lepton operators without the color factor $N_c$.}.
 The solution is
\begin{eqnarray}
  C_{iF}(\Lambda_\chi) &=& 1.03 \ C_{iF}(\Lambda_{\rm EW})
 + 3.0\times 10^{-4} \ {\rm GeV}\, C_{ui}^T(\Lambda_{\rm EW})
-3.2\times 10^{-4} \ {\rm GeV}\, C_{di}^T(\Lambda_{\rm EW})
\nonumber \\
&- &6.4\times 10^{-3}\  {\rm GeV}\, C_{si}^T(\Lambda_{\rm EW})
+0.16\ {\rm GeV}\,C_{ci}^T(\Lambda_{\rm EW})
  - 0.18\ {\rm GeV}\, C_{bi}^T(\Lambda_{\rm EW})
\;.
\end{eqnarray}
We will numerically include the effect of the above renormalization group corrections when running up to the electroweak scale below.
Due to the suppression by the light quark mass, the constraint on the tensor operators through the above mixing effect would be rather weak and thus we will not consider it in the following analysis. Note that, however, it can lead to a constraint on the WCs with heavy quark flavors which is beyond the scope of this work.

\subsection{Matching to the SMNEFT}
\label{sec:SMNEFT}
SMNEFT describes NP which enters at a sufficiently high scale above the electroweak scale. See Appendix~\ref{sec:SMNEFTbasis} for a complete list of SMNEFT operators involving RH neutrinos $N$ up to dim-7 and the relevant dim-6 and dim-7 operators without $N$. LNEFT should be matched to SMNEFT at the electroweak scale $\mu=m_W$ in order to constrain NP. We list the relevant tree-level matching conditions for the LNC and LNV cases in Table~\ref{match}.
Here $v=(\sqrt{2} G_F)^{-1/2}\simeq 246\, \mathrm{GeV}$ is the SM Higgs vacuum expectation value (vev), and $D$ is the unitary matrix transforming left-handed up-type quarks between flavor $u_L^\prime$ and mass eigenstate $u_L$, i.e. $u_L^\prime=D^\dagger u_L$. Under a chosen flavor basis, the flavor and mass eigenstates are identical for the left-handed down-type quarks and RH $u, d$ quarks, and $D$ is then the usual CKM matrix.

\begin{table}[tb]
\centering
\resizebox{\linewidth}{!}{
\renewcommand{\arraystretch}{1.2}
\begin{tabular}{|c| l l |}
\hline
Class & \multicolumn{2}{|c|}{Matching of the Wilson coefficients at the electroweak scale $\Lambda_{\rm EW}$}
\\
\hline
\hline
LNC& $C_{u\nu 1}^{V,pr\alpha\beta}=D_{px}D_{ry}^*\big(C_{l q}^{(1),\alpha\beta xy}+C_{l q}^{(3),\alpha\beta xy}\big)
-\frac{\bar g_Z^2}{M_Z^2} [Z_{u_L}]_{pr} [Z_{\nu}]_{\alpha\beta}$
&
$C_{u\nu 2}^{V,pr\alpha\beta}=C_{l u}^{\alpha\beta pr}-\frac{\bar g_Z^2}{M_Z^2} [Z_{u_R}]_{pr} [Z_{\nu}]_{\alpha\beta}$
\\
$\nu\nu$ case &$C_{d\nu 1}^{V,pr\alpha\beta}=C_{l q}^{(1),\alpha\beta pr}-C_{l q}^{(3),\alpha\beta pr}
-\frac{\bar g_Z^2}{M_Z^2} [Z_{d_L}]_{pr} [Z_{\nu}]_{\alpha\beta}$
& $C_{d\nu 2}^{V,pr\alpha\beta}=C_{l d}^{\alpha\beta pr}
-\frac{\bar g_Z^2}{M_Z^2} [Z_{d_R}]_{pr} [Z_{\nu}]_{\alpha\beta}$
\\
\hline
LNC & $C_{uN 1}^{V,pr\alpha\beta}=D_{px}D_{ry}^*C_{QN}^{xy\alpha\beta}-\frac{\bar g_Z^2}{M_Z^2} [Z_{u_L}]_{pr} [Z_N]_{\alpha\beta}$
&  $C_{uN 2}^{V,pr\alpha\beta}=C_{uN}^{pr\alpha\beta}-\frac{\bar g_Z^2}{M_Z^2} [Z_{u_R}]_{pr} [Z_N]_{\alpha\beta}$
\\
$NN$ case & $C_{dN 1}^{V,pr\alpha\beta}=C_{QN}^{pr\alpha\beta}-\frac{\bar g_Z^2}{M_Z^2} [Z_{d_L}]_{pr} [Z_N]_{\alpha\beta} $
& $C_{dN 2}^{V,pr\alpha\beta}=C_{dN}^{pr\alpha\beta}
-\frac{\bar g_Z^2}{M_Z^2} [Z_{d_R}]_{pr} [Z_N]_{\alpha\beta}$
\\
\hline
& $C_{\nu NF}^{\alpha\beta}~~~=+{v\over \sqrt{2}}e\big(C_{NB}^{\alpha\beta} +C_{NW}^{\alpha\beta}\big)$ &
\\
LNC & $C_{u\nu N1}^{S,pr\alpha\beta}=0$
& $C_{u\nu N2}^{S,pr\alpha\beta}=+D_{rx}^*C_{QuNL}^{xp\beta \alpha*}$
\\
$\nu N$ case & $C_{d\nu N1}^{S,pr\alpha\beta}=+C_{LNQd}^{\alpha\beta pr}-\frac{1}{2}C_{LdQN}^{\alpha rp\beta}$
& $C_{d\nu N2}^{S,pr\alpha\beta}=0$
\\
&$C_{u\nu N}^{T,pr\alpha\beta}=0$
&$C_{d\nu N}^{T,pr\alpha\beta}= -\frac{1}{8}C_{LdQN}^{\alpha rp\beta}$
\\
\hline
\hline
 & $C_{\nu \nu F}^{\alpha\beta}~~~~=+\frac{1}{4}v^2e\big(2C_{LHB}^{\alpha\beta}+C_{LHW}^{\beta\alpha}-C_{LHW}^{\alpha\beta}\big)$  &
\\
LNV & $C_{u\nu 1}^{S,pr\alpha\beta}=0$
& $C_{u\nu 2}^{S,pr\alpha\beta}=+\frac{v}{2\sqrt{2}}D_{px}\big( C_{\bar{Q}uLLH}^{xr\alpha\beta}+C_{\bar{Q}uLLH}^{xr\beta\alpha} \big)$
\\
$\nu\nu$ case & $C_{d\nu 1}^{S,pr\alpha\beta}=-\frac{v}{4\sqrt{2}}\big( C_{\bar{d}LQLH1}^{p\alpha r\beta}+C_{\bar{d}LQLH1}^{p\beta r\alpha} \big)$
& $C_{d\nu 2}^{S,pr\alpha\beta}=0$
\\
&$C_{u\nu}^{T,pr\alpha\beta}=0$
&$ C_{d\nu}^{T,pr\alpha\beta}=+\frac{v}{16\sqrt{2}}\big(C_{\bar{d}LQLH1}^{p\alpha r\beta}-C_{\bar{d}LQLH1}^{p\beta r\alpha}\big)$
\\
\hline
& $C_{NN F}^{\alpha\beta}~~=+\frac{1}{2}v^2e\big(C_{NHB}^{\alpha\beta}-C_{NHW}^{\alpha\beta}\big)$ &
\\
LNV & $C_{uN 1}^{S,pr\alpha\beta}=-{v \over 4\sqrt{2}}D_{px}\big(C_{QNuH}^{x\alpha \beta r}+C_{QNuH}^{x\beta\alpha r}\big)$
& $C_{uN 2}^{S,pr\alpha\beta}=+{v \over \sqrt{2}}D_{rx}^*C_{uQNH}^{px\alpha\beta}$
\\
$N N$ case & $C_{dN 1}^{S,pr\alpha\beta}=-{v \over 4\sqrt{2}}\big(C_{QNdH}^{p\alpha \beta r}+C_{QNdH}^{p\beta\alpha r}\big)$
& $C_{dN 2}^{S,pr\alpha\beta}=+{v \over \sqrt{2}}C_{dQNH}^{pr\alpha\beta}$
\\
& $C_{uN}^{T,pr\alpha\beta}=+{v \over 16\sqrt{2}}D_{px}\big(C_{QNuH}^{x\alpha \beta r}-C_{QNuH}^{x\beta\alpha r}\big)$
& $ C_{dN}^{T,pr\alpha\beta}=+{v \over 16\sqrt{2}}\big(C_{QNdH}^{p\alpha \beta r}-C_{QNdH}^{p\beta\alpha r}\big)$
\\
\hline
LNV & $C_{u\nu N1}^{V,pr\alpha\beta}=-{v \over \sqrt{2}}D_{px}D_{ry}^*C_{QNLH1}^{xy\beta\alpha}
-\frac{\bar g_Z^2}{M_Z^2} [Z_{u_L}]_{pr} [Z_{\nu N}]_{\alpha\beta}$
& $C_{u\nu N2}^{V,pr\alpha\beta}=-{v \over \sqrt{2}}C_{uNLH}^{pr\beta\alpha}
-\frac{\bar g_Z^2}{M_Z^2} [Z_{u_R}]_{pr} [Z_{\nu N}]_{\alpha\beta}$
\\
$\nu N$ case & $C_{d\nu N1}^{V,pr\alpha\beta}=-{v \over \sqrt{2}}\big(C_{QNLH1}^{pr\beta\alpha}-C_{QNLH2}^{pr\beta\alpha}\big)-\frac{\bar g_Z^2}{M_Z^2} [Z_{d_L}]_{pr} [Z_{\nu N}]_{\alpha\beta} $
& $C_{d\nu N2}^{V,pr\alpha\beta}=-{v \over \sqrt{2}}C_{dNLH}^{pr\beta\alpha}
-\frac{\bar g_Z^2}{M_Z^2} [Z_{d_R}]_{pr} [Z_{\nu N}]_{\alpha\beta}$
\\
\hline
\end{tabular}
}
\caption{The matching result of the LNEFT and SMNEFT at the electroweak scale $\Lambda_{\rm EW}$.
The corresponding operators associated with the above SMNEFT Wilson coefficients are collected in Appendix~\ref{sec:SMNEFTbasis}. Note that in LNC $\nu\nu$ case, the notation of the Warsaw basis~\cite{Grzadkowski:2010es} is adopted.}
\label{match}
\end{table}%

Note that SMNEFT operators modify the $Z$ boson couplings from their SM values $[Z_f]_{pr} = \delta_{pr} (T_3-Q s_W^2)$ and the modified couplings are given by~\cite{Jenkins:2017jig}
\begin{align}
	[Z_{\nu}]_{pr} & = \frac12 \delta_{pr} - \frac{v^2}{2}\Big(C_{Hl}^{(1),pr} -C_{Hl}^{(3),pr}\Big)\; ,&
	[Z_{N}]_{pr} & = - \frac{v^2}{2} C_{HN}^{pr}\;, \nonumber \\
	[Z_{e_L}]_{pr} & = \Big(-\frac12 + s_W^2\Big) \delta_{pr} - \frac{v^2}{2}\Big(C_{Hl}^{(1),pr}+C_{Hl}^{(3),pr}\Big) \; , &
	[Z_{e_R}]_{pr}&=s_W^2\delta_{pr}-{v^2\over 2}C^{pr}_{He}\;,  \nonumber \\
	[Z_{u_L}]_{pr} & = \Big(\frac12 -\frac23 s_W^2\Big) \delta_{pr} -\frac{v^2}{2}\Big(C_{Hq}^{(1),pr}-C_{Hq}^{(3),pr}\Big) \; , &
	[Z_{u_R}]_{pr}&=-{2\over 3}s_W^2\delta_{pr}-{v^2\over 2}C^{pr}_{Hu}\;, \nonumber \\
	[Z_{d_L}]_{pr} & = \Big(-\frac12+\frac13 s_W^2\Big) \delta_{pr} - \frac{v^2}{2}\Big(C_{Hq}^{(1),pr}+C_{Hq}^{(3),pr}\Big)\; ,&
	[Z_{d_R}]_{pr}&={1\over 3}s_W^2\delta_{pr}-{v^2\over 2}C^{pr}_{Hd}\;, \nonumber \\
	[Z_{\nu N}]_{pr} & = {v^3\over 4\sqrt{2}}\Big( C^{rp}_{NL1}+2C^{rp}_{NL2}\Big) \; ,
\end{align}
where $[Z_N]_{pr}$ is the modified neutral current coupling to the RH neutrinos defined via ${\cal L}_Z\supset -\bar g_Z [Z_N]_{pr}\bar N_p\slashed{Z}N_r$ and $[Z_{\nu N}]_{pr}$ is the modified neutral current coupling to the LNV current $\overline{\nu^C}\gamma_\mu N$ defined via ${\cal L}_Z\supset -\bar g_Z [Z_{\nu N}]_{pr}\overline{\nu^C_p}\slashed{Z}N_r+{\rm h.c.}$. For simplicity, we do not consider the couplings $[Z_N]$ and $[Z_{\nu N}]$ modified by $C_{HN}$, $C_{NL1}$ and $C_{NL2}$ below. One has $\bar g_Z={e\over s_W c_W}$ with $s_W (c_W)$ being the sine (cosine) of the Weinberg angle as we neglect the contributions from SMEFT operators in the Higgs sector. In the following discussion we also do not consider contributions from the operators $C_{Hl}^{(1)}$, $C_{Hl}^{(3)}$, $C_{Hq}^{(1)}$, $C_{Hq}^{(3)}$, $C_{He}$ and $C_{Hq}$, since they are strongly constrained from electroweak precision measurements.

As only the quark-flavor diagonal $u,d$ or $s$ quark bilinears contribute to the CE$\nu$NS process and the light unflavored meson invisible decays, we will simplify the notation for the LNEFT operators and the corresponding Wilson coefficients by dropping the superscripts for the quark fields and taking the subscript $q$ to be either $u$, or $d$, or $s$ to indicate the specific quark flavor. For instance, ${\cal O}_{u\nu1}^V=(\overline{u_L}\gamma_\mu u_L)(\overline{\nu}\gamma^\mu\nu)$ and the corresponding Wilson coefficient is $C_{u\nu1}^V$ for $q=u$.

\section{Coherent neutrino-nucleus scattering}
\label{sec:CEnuNS}

In the COHERENT experiment, the Spallation Neutron Source produces $\nu_\mu$, $\bar{\nu}_\mu$ and $\nu_e$ from the decay of stopped $\pi^+$ and $\mu^+$. Each neutrino flavor reaches the CsI[Na] detector and contributes to the neutrino flux. The expected number of CE$\nu$NS events depends on the neutrino flux and the CE$\nu$NS differential cross section $d\sigma/dT$ with $T$ being the recoil energy of the nucleus.
The differential cross section for $\overset{\scriptscriptstyle(-)}{\nu} \mathcal{N}\to X \mathcal{N}$  coherent scattering, where $X\in \{\nu,\bar\nu, N, \bar N\}$ denotes a neutrino, is at leading order given by~\cite{Lindner:2016wff,Chang:2020jwl}
\begin{eqnarray}
{d\sigma\over dT}&=&{G_F^2 M\over 4\pi} \Big[\xi_S^2 \frac{T}{T_{\rm max}} + \xi_V^2 \Big(1-{T\over T_{\rm max}}\Big) +\xi_T^2\Big(1-{T\over 2T_{\rm max}}\Big)+e^2A_M^2\Big({1\over MT}-{1\over ME_\nu}\Big)\Big]
\;,
\label{eq:diffeq}
\end{eqnarray}
where $M$ is the nucleus mass, $E_\nu$ is the energy of the incoming neutrino and the maximal value of recoil energy $T$ is $T_{\rm max}={2E_\nu^2\over M+2E_\nu}\simeq {2E_\nu^2\over M}$.
This cross section formula holds for negligible neutrino mass in final states, and thus applies for RH neutrino masses $m_N\lesssim 0.5$ MeV (see e.g. Ref.~\cite{Chang:2020jwl}) irrespective of the mixing between LH and RH neutrinos. The interference terms are suppressed by $T/E_\nu$~\cite{Chang:2020jwl} and are thus not included here.

The $\xi_S, \xi_V, \xi_T$ and $A_M$ constants in Eq.~(\ref{eq:diffeq}) define the effective parameters describing the neutrino-nucleus interactions for scalar, vector, tensor and dipole currents, respectively. They depend on the Wilson coefficients of relevant currents, the number of protons $\mathbb{Z}$ (and neutrons $\mathbb{N}$) in the nucleus, the quantities connecting the quark-level matrix elements and the nucleon-level ones, and the nuclear form factor $F_p$ for protons (and $F_n$ for neutrons). By assuming that one single parameter is present at nuclear level at a time, the constraints on these effective parameters were studied through fitting the COHERENT data. The 90\% CL bounds for the $\xi_S$ and $\xi_T$ parameters are~\cite{AristizabalSierra:2018eqm}
\begin{align}
  \left|\frac{\xi_S}{\mathbb{N} F(Q^2)}\right| & < 0.62\;, 
				    &
  \left|\frac{\xi_T}{\mathbb{N} F(Q^2)}\right| & < 0.591 \;,
\end{align}
where $F(Q^2)$ is the Helm form factor with $Q$ being the transferred energy. The 90\% CL bound on the dipole operators is given by~\cite{Chang:2020jwl}
\begin{eqnarray}
A_M=a_M v \mathbb{Z} F_p(Q^2)\;, \quad {1\over 2}a_M^2\lesssim 7.2\times 10^{-8}\;,
\end{eqnarray}
where the factor of $1/2$ accounts for the missing projection operator in the cross section calculation in Ref.~\cite{Chang:2020jwl}. The scalar, tensor and dipole operators have no interference with the SM neutral current and the above bounds apply to both LNC and LNV cases.

For the vector currents the situation is more complicated and we have to distinguish between LNC and LNV operators. As listed in Table~\ref{match}, there is a  SM contribution to the LNC vector operators with same-flavor quarks and same-flavor active neutrinos
\begin{align}
C_{q\nu1(2),\rm SM}^{V,pr\alpha\beta} & = -\frac{\bar g_Z^2}{2M_Z^2} (T_3-Q_q s_W^2) \delta_{pr} \delta_{\alpha\beta}
\end{align}
in terms of isospin $T_3$ and electric charge $Q_q$. Thus, the interference with the SM has to be taken into account for the NP part of these operators. There is no interference with the SM for the other vector operators.
We thus discuss the constraints separately in the following subsections based on the recent studies of the COHERENT experiment for non-standard interactions (NSIs)\footnote{The relationship between the chiral LNEFT operator basis and NSIs is discussed in Appendix~\ref{sec:NSI}.}~\cite{Giunti:2019xpr} and sterile neutrinos~\cite{Chang:2020jwl}. Both of these studies provide constraints on the quark-level Wilson coefficients.
Next we derive the matrix elements of scattering processes in terms of the LNEFT operators and translate the above bounds to the constraints on the LNEFT Wilson coefficients.

\subsection{LNC case}

The relevant Lagrangian for neutrino-nucleus scattering in the LNC case is given by
\begin{eqnarray}
{\cal L}_{\rm LNC}&\supset&\sum_{\rho,\sigma \atop q=u,d,s}\left[{1\over 2}C_{q\nu1}^{V,\rho\sigma} (\overline{q_L}\gamma_\mu q_L)(\overline{\nu_\rho}\gamma^\mu \nu_\sigma)+
{1\over 2}C_{q\nu2}^{V,\rho\sigma}(\overline{q_R}\gamma_\mu q_R)(\overline{\nu_\rho}\gamma^\mu \nu_\sigma)
+C_{q\nu N1}^{S,\rho\sigma} (\overline{q_L} q_R)(\overline{\nu_\rho} N_\sigma)
\right.\;
\nonumber
\\
&&
\left.+C_{q\nu N2}^{S,\rho\sigma}(\overline{q_R} q_L)(\overline{\nu_\rho} N_\sigma)
+C_{q\nu N}^{T,\rho\sigma} (\overline{q_L} \sigma^{\mu\nu}q_R)(\overline{\nu_\rho}\sigma_{\mu\nu} N_\sigma) + C_{\nu N F}^{\rho\sigma} \bar\nu_\rho \sigma_{\mu\nu} N_\sigma F^{\mu\nu}  \right]+h.c.\;,
\end{eqnarray}
where $\rho, \sigma$ sum over the flavors of active neutrinos $\nu$ and/or RH neutrinos $N$.
We first consider the short-distance contribution induced by the four-fermion operators.
The matrix elements at the nucleus level for the neutrino-nucleus scattering $\nu_\alpha(p_1) {\cal N}(k_1)\to\nu_\beta/N_\beta(p_2) {\cal N}(k_2)$ 
are
\begin{eqnarray}
\mathcal{M}(\nu_\alpha {\cal N}\to \nu_\beta {\cal N})
&=&{1\over 2}C_{{\cal N}\nu}^{V,\alpha\beta*}(\overline{u_{\nu}}\gamma_\mu P_Lu_\nu)\bar {\cal N}\gamma^\mu{\cal N} \;, \nonumber
\\
\mathcal{M}(\nu_\alpha{\cal N}\to N_\beta{\cal N})&=&{1\over 2}C^{S,\alpha\beta*}_{{\cal N}\nu N}(\overline{u_N}P_Lu_\nu)\bar {\cal N}{\cal N}+
C^{T,\alpha\beta*}_{{\cal N}\nu N}(\overline{u_N}\sigma_{\mu\nu}P_Lu_\nu)\bar{\cal N}\sigma^{\mu\nu}{\cal N}\;,
\label{eq:LNCmat}
\end{eqnarray}
where the spin-dependent terms are neglected as they are suppressed by ${\cal O}(E/m_{p/n})$ with respect to spin-independent terms. One should note that the terms with tensor quark current have the property $\sigma_{\mu\nu}P_{L/R}\otimes \sigma^{\mu\nu}P_{L/R}=\sigma_{\mu\nu}P_{L/R}\otimes \sigma^{\mu\nu}$ and thus do not lead to spin-dependent terms.
The matrix elements at the quark level
and the ones for antineutrino nucleus scattering $\bar\nu_\alpha(p_1) {\cal N}(k_1)\to\bar\nu_\beta/\bar N_\beta(p_2) {\cal N}(k_2)$
are given in Appendix~\ref{sec:amplitude}
as reference.
In the above matrix elements for nucleus ``$i$'', the coefficients
\begin{eqnarray}
C^{V,\alpha\beta}_{{\cal N}\nu}&=&\mathbb{Z}_i\left[2(C_{u\nu1}^{V,\alpha\beta}+C_{u\nu2}^{V,\alpha\beta})+C_{d\nu1}^{V,\alpha\beta}+C_{d\nu2}^{V,\alpha\beta}\right]F_p(Q^2)\nonumber \\
&&+\mathbb{N}_i\left[C_{u\nu1}^{V,\alpha\beta}+C_{u\nu2}^{V,\alpha\beta}+2(C_{d\nu1}^{V,\alpha\beta}+C_{d\nu2}^{V,\alpha\beta})\right]F_n(Q^2) \; , \nonumber
\\
C^{S,\alpha\beta}_{{\cal N}\nu N}&=&\sum_{q=u,d,s}(C^{S,\alpha\beta}_{q\nu N1}+C^{S,\alpha\beta}_{q\nu N2})\left[\mathbb{Z}_i{m_p\over m_q} f_{T_q}^p F_p(Q^2)+\mathbb{N}_i{m_n\over m_q} f_{T_q}^n F_n(Q^2)\right]\; , \nonumber
\\
C^{T,\alpha\beta}_{{\cal N}\nu N}&=&\sum_{q=u,d,s}C^{T,\alpha\beta}_{q\nu N}\left[\mathbb{Z}_i\delta_q^p F_p(Q^2) +\mathbb{N}_i\delta_q^n F_n(Q^2)\right]\; ,
\label{LHC:Coeff}
\end{eqnarray}
parameterize the vector, scalar and tensor contributions~\cite{Bischer:2019ttk}. The number of neutrons and protons for Caesium and Iodine are $\mathbb{N}_{\rm Cs}=77.9, \mathbb{Z}_{\rm Cs}=55$ and $\mathbb{N}_{\rm I}=73.9, \mathbb{Z}_{\rm I}=53$, respectively. We assume the proton and neutron form factors are equal to the Helm form factor, i.e. $F_p(Q^2)=F_n(Q^2)=F(Q^2)$. The connections between various quark currents and the nucleon-level ones can be found for instance in Refs.~\cite{DelNobile:2013sia,Bishara:2017pfq}. $f^{p/n}_{T_q}$ and $\delta^{p/n}_q$ are the nucleon form factors for scalar and tensor currents, respectively. For later numerical analysis, we take the following default values from \texttt{micrOMEGAs 5.2}~\cite{Belanger:2008sj,Belanger:2018ccd}
\begin{align}
f^p_{T_u}=&0.0153\;, &
f^p_{T_d}=&0.0191\;, &
f^p_{T_s}=&0.0447\;, \nonumber \\
\delta^p_u=&0.84\;, &
\delta^p_d=&-0.23\;, &
\delta^p_s=&-0.046\;, \nonumber \\
f^n_{T_u}=&0.0110\;, &
f^n_{T_d}=&0.0273\;, &
f^n_{T_s}=&0.0447\;, \nonumber \\
\delta^n_u=&-0.23\;, &
\delta^n_d=&0.84\;, &
\delta^n_s=&-0.046\;.
\end{align}

Using the expressions for the matrix elements in Eq.~\eqref{eq:LNCmat} it is straightforward to compare them with those in Refs.~\cite{Lindner:2016wff,AristizabalSierra:2018eqm} and relate the LNEFT Wilson coefficients to the $\xi$ parameterization\footnote{The relationship between the chiral LNEFT operator basis and the quark-level parameterization in Ref.~\cite{AristizabalSierra:2018eqm} is given in Appendix~\ref{sec:CD}}. We obtain the following constraints on the scalar and tensor coefficients
\begin{eqnarray}
{\xi_S^2\over \mathbb{N}^2F^2}
&=&
\sum_{\beta,i}\Big|{1\over \sqrt{2}G_F} \sum_{q=u,d,s} \left(C^{S,\alpha\beta}_{q\nu N1}+C^{S,\alpha\beta}_{q\nu N2}\right)\left({\mathbb{Z}_i\over \mathbb{N}_i}{m_p\over m_q}f^p_{T_q}+{m_n\over m_q}f^n_{T_q}\right)\Big|^2<0.62^2 \;, \\
{\xi_T^2\over \mathbb{N}^2F^2}
&=&8\sum_{\beta,i}\Big|{\sqrt{2}\over G_F} \sum_{q=u,d,s} C^{T,\alpha\beta}_{q\nu N}\left({\mathbb{Z}_i\over \mathbb{N}_i}\delta_q^p+\delta_q^n\right)\Big|^2<0.591^2 \;.
\end{eqnarray}
These bounds apply for initial state neutrino flavor $\alpha=e$ or $\mu$.
The 90\% CL bounds on the quark-level vector Wilson coefficients can be read off from Fig.~12 in Ref.~\cite{Giunti:2019xpr}.
There is interference between the NP contribution and the SM contribution for LNC vector operators with same-flavor active neutrinos. The interference leads to the following constraints for the NP part of the Wilson coefficient for neutrino flavors $ee$ and $\mu\mu$
\begin{align}
  \frac{C_{u\nu1,\rm NP}^{V,ee} + C_{u\nu2,\rm NP}^{V,ee}}{2\sqrt{2}G_F} & \in [-0.45,0.065]\;, \nonumber
       \\
       \frac{C_{d\nu1,\rm NP}^{V,ee} + C_{d\nu2,\rm NP}^{V,ee}}{2\sqrt{2}G_F}& \in [-0.41,0.060]\;, \nonumber
       \\
  \frac{C_{u\nu1,\rm NP}^{V,\mu\mu} + C_{u\nu2,\rm NP}^{V,\mu\mu}}{2\sqrt{2}G_F} & \in [-0.45,-0.34] \cup [-0.049,0.059]\;, \nonumber
       \\
  \frac{C_{d\nu1,\rm NP}^{V,\mu\mu} + C_{d\nu2,\rm NP}^{V,\mu\mu}}{2\sqrt{2}G_F}& \in [-0.41,-0.31] \cup [-0.044,0.054]\;.
  \label{CV: eeormumu}
\end{align}
Note that the allowed region for the operators $C_{q\nu1(2),\rm NP}^{V,\mu\mu}$ consists of two disjoint pieces. Taking into account the SM contribution, the $\Delta \chi^2$ function for the coefficients of vector operators exhibits two minima as shown in Ref.~\cite{Giunti:2019xpr}. When taking the 90\% CL allowed ranges, we obtain two distinct pieces in the fit results of Eq.~(\ref{CV: eeormumu}). There is no interference for the other LNC vector operators
\begin{align}
  \left|\frac{C_{u\nu1,\rm NP}^{V,e\mu} + C_{u\nu2,\rm NP}^{V,e\mu}}{2\sqrt{2}G_F}\right| &< 0.13\;,
       &
 \left| \frac{C_{d\nu1,\rm NP}^{V,e\mu} + C_{d\nu2,\rm NP}^{V,e\mu}}{2\sqrt{2}G_F}\right|& < 0.11\;,
  \\
  \left|\frac{C_{u\nu1,\rm NP}^{V,e\tau} + C_{u\nu2,\rm NP}^{V,e\tau}}{2\sqrt{2}G_F}\right| &<  0.18\;,
       &
  \left| \frac{C_{d\nu1,\rm NP}^{V,e\tau} + C_{d\nu2,\rm NP}^{V,e\tau}}{2\sqrt{2}G_F}\right|& < 0.17\;,
  \\
  \left|\frac{C_{u\nu1,\rm NP}^{V,\mu\tau} + C_{u\nu2,\rm NP}^{V,\mu\tau}}{2\sqrt{2}G_F}\right| &< 0.16\;,
       &
 \left| \frac{C_{d\nu1,\rm NP}^{V,\mu\tau} + C_{d\nu2,\rm NP}^{V,\mu\tau}}{2\sqrt{2}G_F}\right|& < 0.15\;,
\end{align}
and thus their allowed regions are symmetric around zero.
In the following numerical analysis, we will use the weakest bounds for the Wilson coefficients with diagonal neutrino flavors $ee$ and $\mu\mu$ in Eq.~\eqref{CV: eeormumu} to obtain conservative constraints. Note that, if one adopts other bounds, a stronger limit for the relevant Wilson coefficient and a larger corresponding NP scale will be obtained.

Similarly we derive the constraint on the long-distance contribution to neutrino-nucleus scattering.
It is induced by the dipole operator $\calO_{\nu NF}$ and a virtual photon which mediates the neutrino electromagnetic dipole interaction with the electric charge of the quarks.
The nucleon-level matrix element for $\nu_\alpha (p_1)\mathcal{N}(k_1)\to N_\beta(p_2) \mathcal{N}(k_2)$ scattering is
\begin{eqnarray}
\mathcal{M}(\nu_\alpha {\cal N}\to N_\beta {\cal N})=i{eG_F\over q^2}A_{M\nu NF}^{\alpha\beta*}(\overline{u_N}\sigma_{\mu\nu}P_Lu_\nu)\overline{{\cal N}}\gamma^\mu t^\nu {\cal N}\; ,
\end{eqnarray}
where the transferred 4-momentum is given by $q=p_1-p_2=k_2-k_1$. The corresponding matrix element for $\bar\nu_\alpha \mathcal{N}\to N_\beta \mathcal{N}$ is given in Appendix~\ref{sec:amplitude}.
From the matrix element we determine the dipole operator contribution to the differential scattering cross section
\begin{equation}
A_{M\nu NF}^2 = \mathbb{Z}^2 \sum_\beta \left|\frac{2}{G_F} C_{\nu N F}^{\alpha\beta}\right|^2 F_p^2(Q^2) \;,
\end{equation}
where $\mathbb{Z}$ denotes the number of protons and $F_p^2(Q^2)$ the square of the nuclear form factor for protons as discussed above.
Using the result in Ref.~\cite{Chang:2020jwl}, this translates directly into a constraint on the combination of dipole operator Wilson coefficients $\sum_\beta |C_{\nu NF}^{\alpha\beta}|^2$, where we sum over the final state neutrino flavor
\begin{eqnarray}
{1\over 2}a_{M\nu NF}^2
={1\over 2}\sum_{\beta}\Big|{2\over G_F v}C_{\nu N F}^{\alpha\beta} \Big|^2<7.2\times 10^{-8} \;.
\end{eqnarray}

\subsection{LNV case}
For LNV operators with at least one active neutrino $\nu$, the relevant effective Lagrangian is
\begin{eqnarray}
{\cal L}_{\rm LNV}&\supset&\sum_{\rho,\sigma \atop q=u,d,s}\left[C_{q\nu 1}^{S,\rho\sigma}(\overline{q_R} q_L)(\overline{\nu^C_\rho} \nu_\sigma)+
C_{q\nu 2}^{S,\rho\sigma}(\overline{q_L} q_R)(\overline{\nu^C_\rho} \nu_\sigma)
+C_{q\nu}^{T,\rho\sigma} (\overline{q_R} \sigma^{\mu\nu}q_L)(\overline{\nu^C_\rho}\sigma_{\mu\nu} \nu_\sigma)\;
\right.
\\
\nonumber
&&
\left.+C_{q\nu N1}^{V,\rho\sigma} (\overline{q_L}\gamma_\mu q_L)(\overline{\nu^C_\rho}\gamma^\mu N_\sigma)+
C_{q\nu N2}^{V,\rho\sigma}(\overline{q_R}\gamma_\mu q_R)(\overline{\nu^C_\rho}\gamma^\mu N_\sigma) + C_{\nu\nu F}^{\rho\sigma} (\overline{\nu_\rho^C}\sigma_{\mu\nu}\nu_\sigma) F^{\mu\nu}\right]+h.c.\; .
\end{eqnarray}
The dim-6 interactions in the above lead
to the following matrix elements for neutrino-nucleus scattering $\nu_\alpha (p_1) \mathcal{N} (k_1) \to\bar\nu_\beta/\bar N_\beta (p_2) \mathcal{N}(k_2)$ at the nucleus level
\begin{eqnarray}
\mathcal{M}(\nu_\alpha{\cal N}\to \bar\nu_\beta{\cal N})&=&C^{S,\alpha\beta}_{{\cal N}\nu}(\overline{v^C_{\bar{\nu}}}P_Lu_\nu)\bar {\cal N}{\cal N}
-2C^{T,\alpha\beta}_{{\cal N}\nu}(\overline{v^C_{\bar{\nu}}}\sigma_{\mu\nu}P_Lu_\nu)\bar{\cal N}\sigma^{\mu\nu}{\cal N} \; , \nonumber\\
\mathcal{M}(\nu_\alpha {\cal N}\to \bar N_\beta {\cal N})
&=&-{1\over 2}C_{{\cal N}\nu N}^{V,\alpha\beta}(\overline{v^C_{\bar N}}\gamma_\mu P_Lu_\nu)\bar {\cal N}\gamma^\mu{\cal N} \;,
\end{eqnarray}
where the coefficients $C^{S,\alpha\beta}_{{\cal N}\nu}, C^{V,\alpha\beta}_{{\cal N}\nu}$ and $C^{T,\alpha\beta}_{{\cal N}\nu}$ have similar expressions as the LNC case in Eq.~\eqref{LHC:Coeff} with a proper replacement of the quark level Wilson coefficients: $C_{q\nu 1(2)}^V$, $C_{q\nu N1(2)}^S$, and $C_{q\nu N}^T$ by $C_{q\nu N 1(2)}^V$, $C_{q\nu 1(2)}^S$, and $C_{q\nu}^T$, respectively.
The quark-level matrix elements and the matrix elements for antineutrino nucleus scattering $\bar \nu_\alpha \mathcal{N}\to \nu_\beta/N_\beta \mathcal{N}$ are also given in Appendix~\ref{sec:amplitude}.
For the scalar and tensor coefficients we again relate them to the $\xi$ parameterization and get the following constraints
\begin{eqnarray}
{\xi_S^2\over \mathbb{N}^2F^2}
&=&\sum_{\beta,i}\Big|{\sqrt{2}\over G_F} \sum_{q=u,d,s} (C^{S,\alpha\beta}_{q\nu 1}+C^{S,\alpha\beta}_{q\nu 2})({\mathbb{Z}_i\over \mathbb{N}_i}{m_p\over m_q}f^p_{T_q}+{m_n\over m_q}f^n_{T_q})\Big|^2<0.62^2 \;, \\
{\xi_T^2\over \mathbb{N}^2F^2}
&=&8\sum_{\beta,i}\Big|{2\sqrt{2}\over G_F} \sum_{q=u,d,s} C^{T,\alpha\beta}_{q\nu}({\mathbb{Z}_i\over \mathbb{N}_i}\delta_q^p+\delta_q^n)\Big|^2<0.591^2 \;,
\end{eqnarray}
The above constraints apply for $\alpha=e, \mu$ for initial neutrino flavors. The LNV vector currents lead to RH neutrinos in scattering final states and such process has been studied in Refs.~\cite{Brdar:2018qqj,Chang:2020jwl} through fitting the COHERENT data for the $\nu\mathcal{N}\to \chi\mathcal{N}$ scattering. After comparing the amplitudes and translating the bound developed in Ref.~\cite{Chang:2020jwl}, we find the following constraint on the LNV vector Wilson coefficients
\begin{eqnarray}
{1\over 2}\sum_{\beta}\Big|{1\over \sqrt{2}G_F}(C_{q\nu N1}^{V,\alpha\beta}+C_{q\nu N2}^{V,\alpha\beta}) \Big|^2 < 1.1\times 10^{-2}\; ,
\end{eqnarray}
where $\alpha=e, \mu$ again and the factor of $1/2$ accounts for the missing projection operator in the cross section calculation of Ref.~\cite{Chang:2020jwl}.

Finally, the coupling of the photon to active neutrinos via the LNV dipole operator $\mathcal{O}_{\nu\nu F}$ also induces a long-distance contribution to neutrino-nucleus scattering $\nu_\alpha(p_1) \mathcal{N}(k_1)\to \bar \nu_\beta(p_2) \mathcal{N}(k_2)$. The corresponding matrix element is given by
\begin{eqnarray}
\mathcal{M}(\nu_\alpha {\cal N}\to \bar\nu_\beta {\cal N})&=&
-i{eG_F\over q^2}A_{M\nu \nu F}^{\alpha\beta}(\overline{v_{\bar \nu}^C}\sigma_{\mu\nu}P_Lu_\nu)\overline{{\cal N}}\gamma^\mu t^\nu {\cal N}\; ,
\end{eqnarray}
where $q=p_1-p_2=k_2-k_1$ denotes the transferred 4-momentum. The matrix element for the antineutrino-nucleus scattering $\bar \nu_\alpha \mathcal{N}\to \nu_\beta \mathcal{N}$ is given in Appendix~\ref{sec:amplitude}. Both neutrino-nucleus and antineutrino-nucleus scattering are described by the same dipole operator contribution
\begin{equation}
 A_{M\nu\nu F}^2 = \mathbb{Z}^2 \sum_\beta \left|\frac{4}{G_F} C_{\nu \nu F}^{\alpha\beta}\right|^2 F_p^2(Q^2)\;,
\end{equation}
to the differential scattering cross section in Eq.~\eqref{eq:diffeq}.
The constraint given in Ref.~\cite{Chang:2020jwl} allows to place a constraint on $\sum_\beta |C_{\nu\nu F}^{\alpha\beta}|^2$
\begin{eqnarray}
{1\over 2}a_{M\nu\nu F}^2
={1\over 2}\sum_{\beta}\Big|{4\over G_F v}C_{\nu \nu F}^{\alpha\beta} \Big|^2<7.2\times 10^{-8} \; ,
\end{eqnarray}
where $\alpha$ denotes the initial state neutrino flavor. Note, that there is no constraint on the Wilson coefficient of the LNV dipole operator $\mathcal{O}_{NNF}$ from neutrino-nucleus scattering, because the initial state in the COHERENT experiment is always an active neutrino $\nu_\alpha$ or antineutrino $\bar \nu_\alpha$.

\section{Meson invisible decay}
\label{sec:inv}
Next we consider the constraints on the LNEFT Wilson coefficients from meson invisible decays as listed in Table~\ref{tab:meson}. For simplicity we focus on the case, where the mixing between LH and RH neutrinos can be neglected ($|\sin\theta|^2 \lesssim 0.01$).

\begin{table}
\centering
\begin{tabular}{|c|c||c|c|}
\hline
Pseudoscalar meson & Upper limit on BR & Vector meson & Upper limit on BR
\\\hline
$\pi\to$ invisible & $2.7\times 10^{-7}$ & $\omega(782)\to$ invisible & $7.0\times 10^{-5}$
\\
$\eta\to$ invisible & $1.0\times 10^{-4}$ & $\phi(1020)\to$ invisible & $1.7\times 10^{-4}$
\\
$\eta^\prime(958)\to$ invisible &  $5.0\times 10^{-4}$ & &
\\\hline
\end{tabular}
\caption{Relevant constraints on the invisible decays of pseudoscalar mesons $J^{PC}=0^{-+}$ and vector mesons $J^{PC}=1^{--}$~\cite{Tanabashi:2018oca}.}
\label{tab:meson}
\end{table}

\subsection{Light pseudoscalar meson decays}
For a pseudoscalar meson $P$, the transition matrix element to the vacuum state from the scalar, vector, and tensor quark currents are zero. The only non-vanishing matrix elements are for pseudo-scalar currents, axial-vector currents and the anomaly matrix elements. They can be parameterized by the form factors $f_P^q$, $h_P^q$ and $a_P$~\cite{Beneke:2002jn,Cheng:2013fba}
\begin{align}
\left\langle0|\bar q\gamma^\mu\gamma_5q|P(p)\right\rangle=&i f^q_P p^\mu\;, &
\left\langle0|\bar q\gamma_5q|P(p)\right\rangle=&-i {h_P^q\over 2m_q} \;, &
\langle 0| \frac{\alpha_s}{4\pi} G_{a,\mu\nu}\tilde G_a^{\mu\nu} |P(p)\rangle &= a_P\;,
\end{align}
where $\tilde G_{\mu\nu} = \frac12 \epsilon_{\mu\nu\rho\sigma} G^{\rho\sigma}$ and $\epsilon_{0123}=1$ and the form factors satisfy
    $h_P^q =   m_P^2 f_P^q - a_P $.
The form factors for the mesons $\pi^0,\eta, \eta^\prime$ can be expressed in terms of the input form factors $f_q=1.07 f_\pi$, $f_s=1.34f_\pi$, and  $f_\pi=130.2$ MeV~\cite{Tanabashi:2018oca}
\begin{align}
f^u_\pi=&-f^d_\pi={1\over \sqrt{2}}f_\pi \;, &
f^s_\pi=&0  \;; &
h^i_\pi=&m_\pi^2f^i_\pi, \quad i=u,d,s\;,\nonumber
\\
f^u_\eta=&f^d_\eta={c_\phi\over \sqrt{2}}f_q \;, &
f^s_\eta=&-s_\phi f_s  \;;&
h^u_\eta=&h^d_\eta={c_\phi\over \sqrt{2}}h_q \;, &
h^s_\eta=&-s_\phi h_s  \;,\nonumber
\\
f^u_{\eta^\prime}=&f^d_{\eta^\prime}={s_\phi\over \sqrt{2}}f_q  \;, &
f^s_{\eta^\prime}=&c_\phi f_s\;; &
h^u_{\eta^\prime}=&h^d_{\eta^\prime}={s_\phi\over \sqrt{2}}h_q  \;, &
h^s_{\eta^\prime}=&c_\phi h_s\;, &
\end{align}
where $s_\phi=\sin\phi$ and $c_\phi=\cos\phi$ with $\phi=39.3^\circ$ being the mixing angle between flavor $SU(3)$ octet $\eta_8$ and singlet $\eta_1$. We assume isospin symmetry following the FKS scheme~\cite{Feldmann:1998vh,Feldmann:1998sh,Feldmann:1999uf} for the form factors of $\eta$ and $\eta^\prime$ and take the numerical values from Ref.~\cite{Beneke:2002jn} unless otherwise stated. The pseudoscalar input form factor $h_q$ and $h_s$ can be expressed in terms of $f_q$, $f_s$ and $\phi$ as follows
\begin{align}
h_q =& f_q (m_\eta^2c^2_\phi + m_{\eta^\prime}^2 s^2_\phi)-\sqrt{2}f_s (m_{\eta^\prime}^2-m_{\eta}^2)s_\phi c_\phi \; ,\nonumber
\\
h_s=&f_s (m_{\eta^\prime}^2 c^2_\phi + m_{\eta}^2 s^2_\phi)-{f_q\over \sqrt{2}} (m_{\eta^\prime}^2-m_{\eta}^2)s_\phi c_\phi \; .
\end{align}
Given the definition of the above-listed form factors, we can write the branching ratio for the invisible decay of a pseudoscalar meson to neutrinos as\footnote{Calculational details are collected in Appendix~\ref{sec:invdecay} }
\begin{align}
  \label{eq:BrPinv}
{\cal B}(P\to {\rm inv.})
=\frac{\tau_{P}m_P}{16\pi }\sum_{\alpha,\beta}\Bigg\{
&
2\left|\frac{m_Nf_P^q}{2}\left( C_{qN 1}^{V,\alpha\beta}-C_{qN 2}^{V,\alpha\beta}\right)\right|^2\left(1-4\frac{m_N^2}{m_P^2}\right)^{1\over2}
\\\nonumber&
+  2\left| \frac{h_P^q}{4 m_q}  \left(C_{q\nu N1}^{S,\alpha\beta}-C_{q\nu N2}^{S,\alpha\beta}\right)\right|^2 \left(1-\frac{m_N^2}{m_P^2}\right)^2
\\\nonumber&
+ \left|\frac{h_P^q}{2m_q}\left( C_{q\nu 1}^{S,\alpha\beta}-C_{q\nu 2}^{S,\alpha\beta}\right)\right|^2
\\\nonumber&
+\left| \frac{h_P^q}{2m_q}\left( C_{qN1}^{S,\alpha\beta}-C_{qN 2}^{S,\alpha\beta}\right)\right|^2 \left(1-2\frac{m_N^2}{m_P^2}\right)\left(1-4 \frac{m_N^2}{m_P^2}\right)^{1\over2}
\\\nonumber&
+2\left|m_N f_P^q\left( C_{q\nu N 1}^{V,\alpha\beta}-C_{q\nu N 2}^{V,\alpha\beta}\right)\right|^2\left(1-\frac{m_N^2}{m_P^2}\right)^2
\Bigg\}\;,
\end{align}
where we implicitly sum over light quark flavors $u,d,s$.
The contributions in the first two lines describe LNC decays and the remaining lines LNV decays. In the SM, there is only a contribution to the operators $\calO_{q\nu1(2)}^V$, whose contribution to the decay is helicity suppressed and thus negligible due to the tiny neutrino masses.

\subsection{Light vector meson decays}
The non-vanishing hadronic matrix element for an unflavored vector meson $V$ with momentum $p$ and polarization vector $\epsilon^\mu_V$ can be parameterized as~\cite{Ball:2006eu,Cheng:2013fba}
\begin{align}
\langle 0|\bar{q}\gamma^\mu q|V(p)\rangle =& f_V^q m_V \epsilon^\mu_V \; , &
\langle 0|\bar{q}\sigma^{\mu\nu} q|V(p)\rangle =& if_V^{T,q} \left(\epsilon^\mu_V p^\nu - \epsilon^\nu_Vp^\mu \right) \; .
\end{align}
In particular the form factors for the vector mesons $\omega\sim {u\bar{u}+d\bar{d}\over \sqrt{2}}$ and $\phi\sim s\bar{s}$ are
\begin{align}
f_\omega^u=f_\omega^d=&{1\over \sqrt{2} }f_\omega\;,  &
f_\omega^s=&0\;, &
f_\omega^{T,u}=f_\omega^{T,d}=&{1\over \sqrt{2} } f_\omega^T\;, &
f_\omega^{T,s}=&0\;,\nonumber
\\
~ f_\phi^{u}=f_\phi^{d}=&0\;, &
f_\phi^s=&f_\phi\;, &
f_\phi^{T,u}=f_\phi^{T,d}=&0\;, &
f_\phi^{T,s}=&f_\phi^{T}\;,
\end{align}
with $f_\omega=187$ MeV, $f_\omega^T=151$ MeV~\cite{Ball:2006eu}, $f_\phi=233~\rm MeV$ and $f_\phi^{T}=177~\rm MeV$~\cite{Becirevic:2003pn}.

Using the definition of these form factors, it is straightforward to derive an expression for the branching ratio of the vector meson invisible decay to neutrinos\footnotemark[\value{footnote}]
\begin{align}
  \label{eq:BrVinv}
{\cal B}(V\to {\rm inv.})
=\frac{\tau_{V}m_V^3}{48\pi }\sum_{\alpha,\beta}\Bigg\{
&
2\left|\frac{f_V^q}{2}\left( C_{q\nu 1, {\rm NP}}^{V,\alpha\beta}+C_{q\nu 2, {\rm NP}}^{V,\alpha\beta}\right)+C^V_{\rm SM}\delta_{\alpha\beta}\right|^2
\\\nonumber
&+2\left|\frac{f_V^q}{2}\left( C_{qN1}^{V,\alpha\beta}+C_{qN 2}^{V,\alpha\beta}\right)\right|^2
\left(1-\frac{m_N^2}{m_V^2}\right)\left(1-4\frac{m_N^2}{m_V^2}\right)^{1\over 2}
\\\nonumber
&+8\left|f_V^{T,q}C_{q\nu N}^{T,\alpha\beta} -eQ_q{f_V^q\over m_V}C_{\nu NF}^{\alpha\beta}\right|^2
\left(1+{m_N^2 \over m_V^2}-2{m_N^4 \over m_V^4}\right)\left(1-\frac{m_N^2}{m_V^2}\right)
\\\nonumber
&+16\left|f_V^{T,q}C_{q\nu}^{T,\alpha\beta}-eQ_q{f_V^q\over m_V}C_{\nu\nu F}^{\alpha\beta}\right|^2
\\\nonumber
&+16\left|f_V^{T,q}C_{qN}^{T,\alpha\beta}-eQ_q{f_V^q\over m_V}C_{NN F}^{\alpha\beta}\right|^2
\left(1+2{m_N^2\over m_V^2}\right)\left(1-4{m_N^2\over m_V^2}\right)^{1\over2}
\\\nonumber
&+4\left|{f_V^q\over 2}\left(C_{q\nu N1}^{V,\alpha\beta}+ C_{q\nu N2}^{V,\alpha\beta}\right)\right|^2
\left(1-{m_N^2 \over 2m_V^2}-{m_N^4 \over 2m_V^4}\right)\left(1-{m_N^2\over m_V^2}\right)
\Bigg\}\;,
\end{align}
where we implicitly sum over light quark flavors $u,d,s$.
The first three lines describe LNC decays and the latter three LNV decays.
The dipole operators contribute to the vector meson invisible decays through a photon propagator and a QED vertex. Here we have split the contribution from the vector operators $C_{q\nu 1(2)}^V$ into the NP contribution and the SM part as
\begin{align}
C^V_{\rm SM}=-{g_Z^2\over4m_Z^2}\Big[\Big({1\over2}-{4\over3}s_W^2\Big)f_V^u-\Big({1\over2}-{2\over3}s_W^2\Big)\left(f_V^d+f_V^s\right)\Big]\;,
\end{align}
with
\begin{align}
C^\omega_{\rm SM}&={g_Z^2s_W^2f_\omega\over6\sqrt{2}m_Z^2}\;, &
C^\phi_{\rm SM}&=\Big({1\over2}-{2\over3}s_W^2\Big){g_Z^2f_\phi\over4m_Z^2}\;.
\end{align}
The SM predictions for the vector meson invisible decays are
\begin{eqnarray}
{\cal B}(\omega\to {\rm inv.})\approx1.5\times 10^{-13}\;, \quad {\cal B}(\phi\to {\rm inv.})\approx3.4\times 10^{-10},
\end{eqnarray}
and consequently negligible compared with the current experimental upper limits listed in Table~\ref{tab:meson}.

Generally, the scalar-type and tensor-type LNEFT operators can only be constrained by pseudoscalar and vector meson invisible decays, respectively. For vector-type operators, both pseudoscalar and vector meson decays are sensitive to LNC $\mathcal{O}^V_{qN1(2)}$ and LNV $\mathcal{O}^V_{q\nu N1(2)}$. The LNC operators $\mathcal{O}^V_{q\nu 1(2)}$ and all dipole operators only contribute to vector meson decay.

\section{Numerical results}
\label{sec:Num}

\begin{figure}[htb!]
\begin{center}
\includegraphics[scale=1,width=0.7\linewidth]{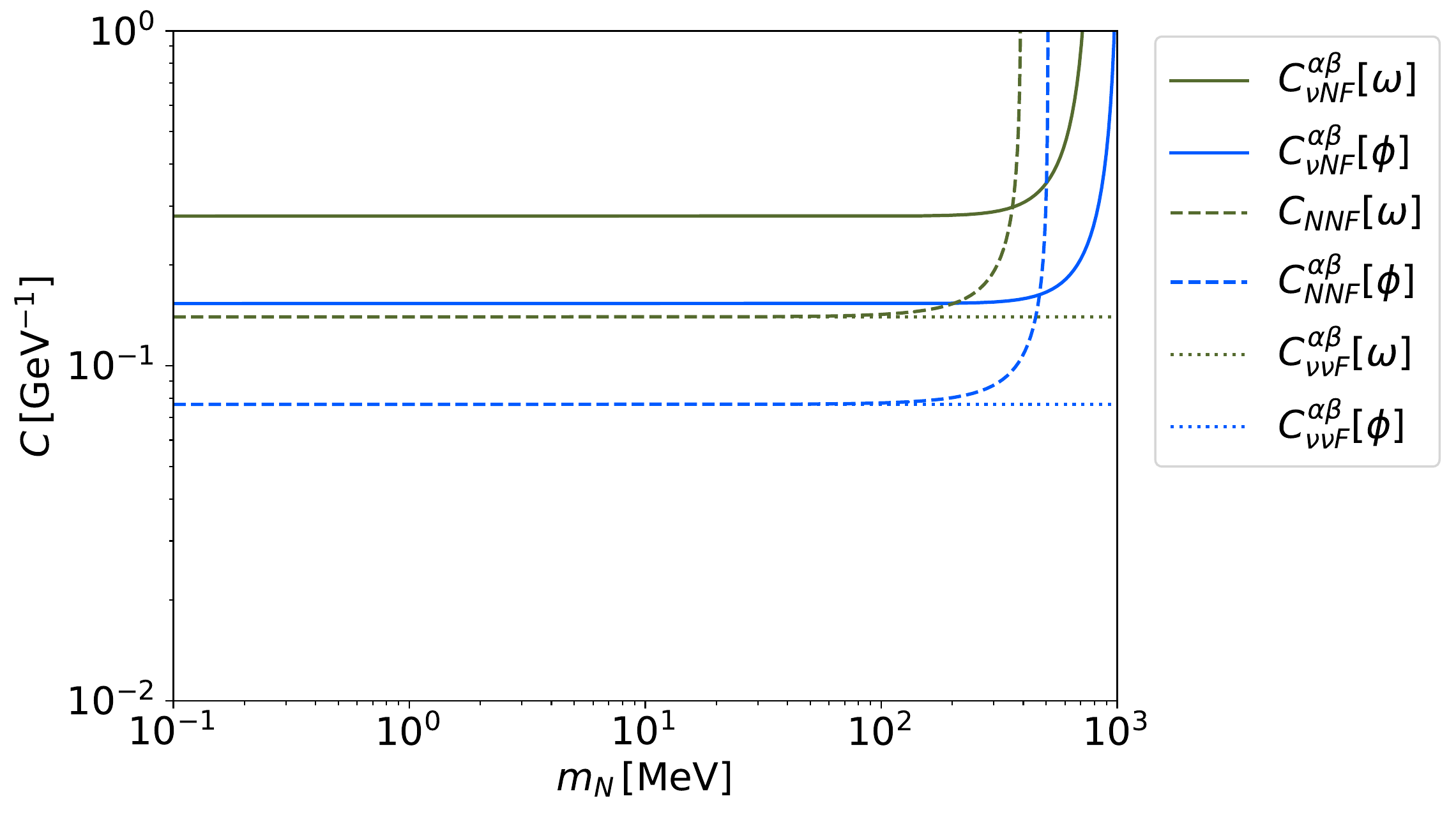}
\end{center}
\caption{
The upper bounds on the LNEFT Wilson coefficients of dipole operators from meson invisible decays as a function of the RH neutrino mass $m_N$.
}
\label{fig:Dipole}
\end{figure}

\begin{figure}[htb!]
\setlength\abovecaptionskip{-12pt}
\begin{center}
\includegraphics[scale=1,width=0.497\linewidth]{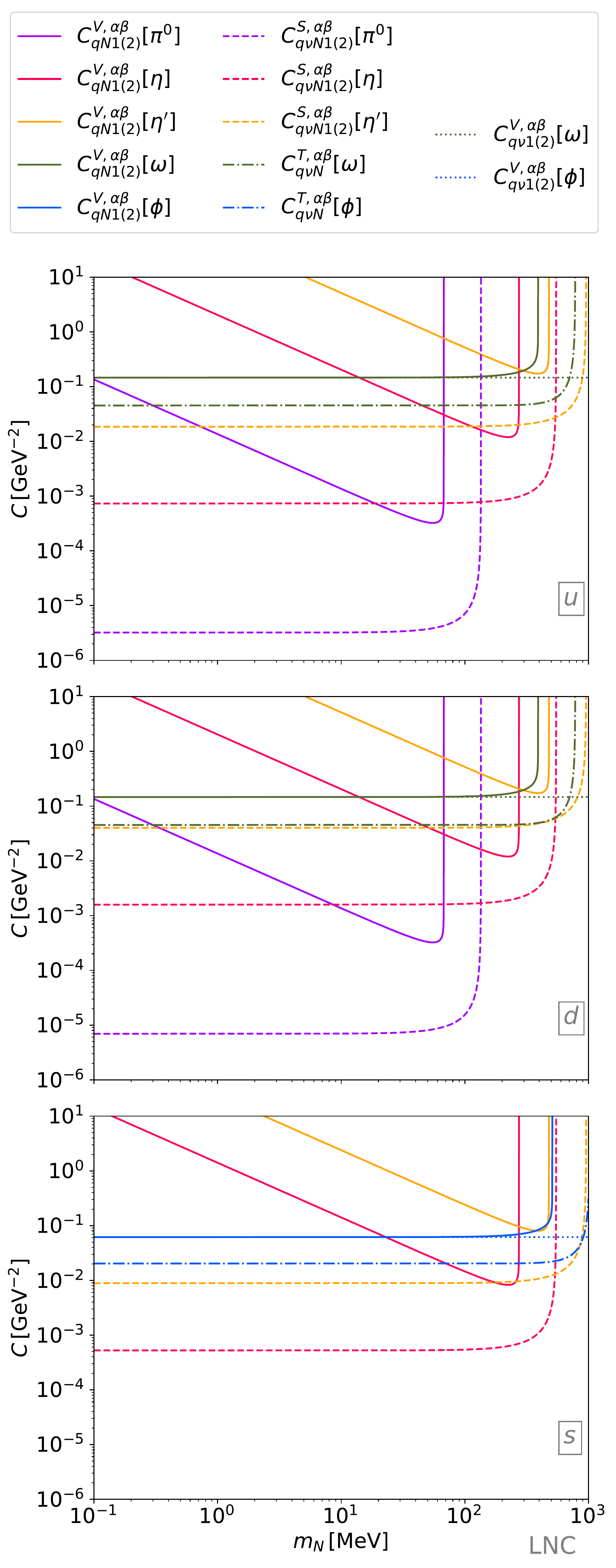}
\includegraphics[scale=1,width=0.497\linewidth]{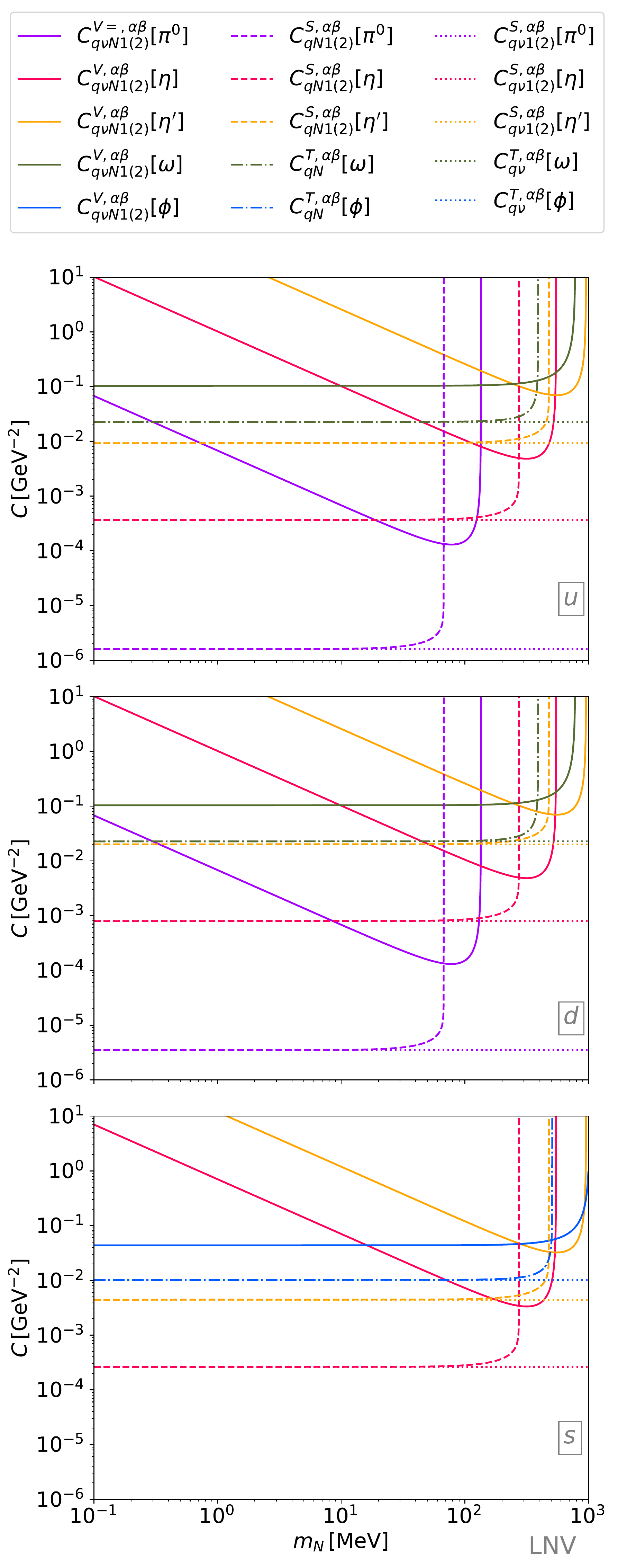}
\end{center}
\caption{
 Upper bounds on the LNEFT Wilson coefficients of dim-6 neutrino-quark operators from meson invisible decays as a function of  $m_N$ for LNC (LNV) operators on the left (right). The top (middle) [bottom] row show WCs for up (down) [strange] quarks.
For the LNV WCs $C_{q\nu 1(2)}^{S,\alpha\beta}$, $C_{qN1(2)}^{S,\alpha\beta}$ we display the components with $\alpha\neq \beta$. The bounds for the corresponding WCs with $\alpha=\beta$ are stronger by a factor $\sqrt{2}$.
}
\label{fig:dim6}
\end{figure}

In this section we present the numerical constraints on the Wilson coefficients of LNEFT and SMNEFT from the CE$\nu$NS process and meson invisible decays. We assume that one operator dominates at a time.
We first show the upper bounds on the LNEFT Wilson coefficients from meson invisible decays as a function of $m_N$ in Figs.~\ref{fig:Dipole} and \ref{fig:dim6}.
The different colored lines correspond to different mesons: $\pi^0$ (purple), $\eta$ (red), $\eta^\prime$ (orange), $\omega$ (dark green) and $\phi$ (blue).

Fig.~\ref{fig:Dipole} shows the constraints for the dipole operators. Solid (dashed) [dotted] lines correspond to the Wilson coefficients $C_{\nu NF}^{\alpha\beta}$ ($C_{NNF}^{\alpha\beta}$) [$C_{\nu\nu F}^{\alpha\beta}$]. The constraints on $C_{NNF}^{\alpha\beta}$ are cut off for smaller RH neutrino masses compared to the ones for $C_{\nu NF}^{\alpha\beta}$ due to the smaller phase space with two massive RH neutrinos in the final state.

In Fig.~\ref{fig:dim6},
solid (dashed) [dot-dashed] lines indicate vector (scalar) [tensor] Wilson coefficients. The horizontal dotted lines show the bounds on the Wilson coefficients without RH neutrino field for completeness.
For $C_{q\nu 1(2)}^{S,\alpha\beta}$ and $C_{qN1(2)}^{S,\alpha\beta}$ with
symmetric neutrino flavors in the LNV case, which are shown in the bottom,
we show
the components with different flavors ($\alpha\neq \beta$). The bounds on the
Wilson coefficients with identical flavors ($\alpha=\beta$) are enhanced by a
factor of $\sqrt{2}$ with respect to those with $\alpha\neq \beta$.
One can see that, from pseudoscalar meson decay, the upper limits on the
$C^V_{qN1(2)}$ in LNC case and the $C^V_{q\nu N1(2)}$ in LNV case both scale
as $\sim 1/m_N$ and are thus less stringent than the constraints from vector
meson decay in the small $m_N$ limit. The bounds on other coefficients turn out to
be a constant if the decay is kinematically allowed.

\begin{table}[htbp!]
\resizebox{\linewidth}{!}{
\renewcommand{\arraystretch}{1.14}
\begin{tabular}{|c|c|c|c|c|c|c|c|}
\hline
LNEFT WC &~~~CE$\nu$NS~~~&
$~\pi^0\to {\rm inv.}~$ & $~\eta\to {\rm inv.}~$ & $~\eta^\prime\to {\rm inv.}~$ & $~\omega \to {\rm inv.}~$ & $~\phi\to {\rm inv.}~$ & $\Lambda_{\rm LNEFT}=\left|C_i\right|^{1\over 4-d}$
\\
$[{\rm GeV}^{4-d}]$ &~ $\alpha=e$ or $\mu$ ~&~$2.7\times10^{-7}$~&~ $1.0\times10^{-4}$ ~&~ $5.0\times10^{-4}$~ &~ $7.0\times10^{-5}$ ~&
$ 1.7\times10^{-4}$ & [GeV]
\\\hline\hline
$C_{u\nu1(2),{\rm NP}}^{V,ee}$ &\cellcolor{gray!35}$1.5\times10^{-5}$ & - & - & - & $1.5\times10^{-1}$ & - & 260
\\\hline
$C_{u\nu1(2),{\rm NP}}^{V,e\mu}$ &\cellcolor{gray!35}$4.3\times10^{-6}$ & - & - & - & $1.5\times10^{-1}$ & - & 480
\\\hline
$C_{u\nu1(2),{\rm NP}}^{V,e\tau}$ &\cellcolor{gray!35} $5.9\times10^{-6}$ & - & - & - & $1.5\times10^{-1}$ & - &410
\\\hline
$C_{u\nu1(2),{\rm NP}}^{V,\mu\mu}$ &\cellcolor{gray!35} $1.5\times10^{-5}$ & - & - & - & $1.5\times10^{-1}$ & - & 260
\\\hline
$C_{u\nu1(2),{\rm NP}}^{V,\mu\tau}$&\cellcolor{gray!35} $5.3\times10^{-6}$ & - & - & - & $1.5\times10^{-1}$ & - & 440
\\\hline
$C_{u\nu1(2),{\rm NP}}^{V,\tau\tau}$&- & - & - & - &\cellcolor{gray!35} $1.5\times10^{-1}$& - & 2.6
\\\hline
$C_{d\nu1(2),{\rm NP}}^{V,ee}$  &\cellcolor{gray!35}$ 1.4\times10^{-5}$  &- & - & - & $1.5\times10^{-1}$  & - & 270
\\\hline
$C_{d\nu1(2),{\rm NP}}^{V,e\mu}$&\cellcolor{gray!35}$ 3.6\times10^{-6}$ &- & - & - & $1.5\times10^{-1}$  & - &  520
\\\hline
$C_{d\nu1(2),{\rm NP}}^{V,e\tau}$&\cellcolor{gray!35}   $ 5.6\times10^{-6}$ &- & - & - & $1.5\times10^{-1}$  & - &420
\\\hline
$C_{d\nu1(2),{\rm NP}}^{V,\mu\mu}$&\cellcolor{gray!35}  $1.4\times10^{-5}$  &- & - & - & $1.5\times10^{-1}$  & - &  270
\\\hline
$C_{d\nu1(2),{\rm NP}}^{V,\mu\tau}$&\cellcolor{gray!35}  $5.0\times10^{-6}$  &- & - & - & $1.5\times10^{-1}$  & - &  450
\\\hline
$C_{d\nu1(2),{\rm NP}}^{V,\tau\tau}$&- &- & - & - & $\cellcolor{gray!35}1.5\times10^{-1}$  & - & 2.6
\\\hline
$C_{s\nu1(2),{\rm NP}}^{V,\alpha\beta}$  & - & - & - & - & - &\cellcolor{gray!35} $6.2\times10^{-2}$ & 4.0
\\\hline\hline 
$C_{uN1(2)}^{V,\alpha\beta}$  & - & - & - & - &\cellcolor{gray!35} $1.5\times10^{-1}$   & - & 2.6
\\\hline
$C_{dN1(2)}^{V,\alpha\beta}$  & - & - & - & - &\cellcolor{gray!35}  $1.5\times10^{-1}$  & - & 2.6
\\\hline
$C_{sN1(2)}^{V,\alpha\beta}$   & - & - & - & - & - &\cellcolor{gray!35}  $6.2\times10^{-2}$ & 4.0
\\\hline\hline 
$C_{\nu NF}^{\alpha\beta}$  &\cellcolor{gray!35} $5.4\times10^{-7}$& - & - & -  &\cellcolor{gray!15}$2.8\times10^{-1}$  & $1.6\times10^{-1}$ & $1.9\times10^{6}$~(1.9)
\\\hline
$C_{u\nu N1(2)}^{S,\alpha\beta}$  &\cellcolor{gray!35} $7.6\times10^{-7}$ &\cellcolor{gray!15}$3.2\times10^{-6}$  & $7.3\times10^{-4}$  & $1.9\times10^{-2}$ &- & - & 1100~(560)
\\\hline
$C_{d\nu N1(2)}^{S,\alpha\beta}$  &\cellcolor{gray!35} $8.8\times10^{-7}$  &\cellcolor{gray!15}$6.9\times10^{-6}$  & $1.6\times10^{-3}$  &$4.0\times10^{-2}$  &- & - & 1100~(380)
\\\hline
$C_{s\nu N1(2)}^{S,\alpha\beta}$  &\cellcolor{gray!35} $9.4\times10^{-6}$ & - &\cellcolor{gray!15} $5.2\times10^{-4}$  &$8.9\times10^{-3}$   &- & - & 330~(44)
\\\hline
$C_{u\nu N}^{T,\alpha\beta}$ &\cellcolor{gray!35} $3.3\times10^{-6}$ & - & - & - &\cellcolor{gray!15} $4.5\times10^{-2}$  & -& 550~(4.7)
\\\hline
$C_{d\nu N}^{T,\alpha\beta}$ &\cellcolor{gray!35}$1.8\times10^{-6}$ & - & - & - &\cellcolor{gray!15}  $4.5\times10^{-2}$  & - & 750~(4.7)
\\\hline
$C_{s\nu N}^{T,\alpha\beta}$   &\cellcolor{gray!35} $1.5\times10^{-5}$ & - & - & - & - &\cellcolor{gray!15}  $4.1\times10^{-2}$ & 250~(4.9)
\\
\hline
\end{tabular}
}
\caption{Constraints on the Wilson coefficients of the LNC operators in the LNEFT. The neutrino flavors for the vector type operators $C_{u(d)\nu 1(2)}^V$ are displayed explicitly. For CE$\nu$NS the initial neutrino flavor is $\alpha=e,\mu$. In other cases the neutrino flavors $\alpha,\beta$ are arbitrary.
The gray cell displays the strongest constraint for each WC. In the last column we also show the effective scale derived from the strongest constraint for each WC. Note that in the last sector, the gray and light gray cells are for $\alpha=e,\mu$ and $\alpha=\tau$ flavors respectively. For the $\alpha=\tau$ case, the effective scale is shown in parentheses (\dots) in the last column.
}
\label{tab:LNEFTLNC}
\end{table}

\begin{table}[h!]
\resizebox{\linewidth}{!}{
\renewcommand{\arraystretch}{1.14}
\begin{tabular}{|c|c|c|c|c|c|c|c|}
\hline
LNEFT WC &~~~CE$\nu$NS~~~&
$\pi^0\to {\rm inv.}$ & $\eta\to {\rm inv.}$ & $\eta^\prime\to {\rm inv.}$ & $\omega \to {\rm inv.}$ & $\phi\to {\rm inv.}$ & $\Lambda_{\rm LNEFT}=\left|C_i\right|^{1\over 4-d}$
\\
$[{\rm GeV}^{4-d}]$ & $\alpha=e$ or $\mu$ &  $2.7\times10^{-7}$ & $1.0\times10^{-4}$ & $5.0\times10^{-4}$ & $7.0\times10^{-5}$ & $1.7\times10^{-4}$ & [GeV]
\\\hline\hline
$C_{\nu\nu F}^{\alpha\beta}$  &\cellcolor{gray!35}  $2.7\times10^{-7}$ & - & - & - & $1.4\times10^{-1}$  & $7.8\times10^{-2}$ & $3.7\times10^6$
\\\hline
$C_{u\nu1(2)}^{S,\alpha\beta}$  &\cellcolor{gray!35}  $3.8\times10^{-7}$ &\cellcolor{gray!15}  $1.6[2.3]\times10^{-6}$  & $3.7[5.2]\times10^{-4}$ & $9.3[13~]\times10^{-3}$  & - & - & 1600~(660)
\\\hline
$C_{d\nu1(2)}^{S,\alpha\beta}$  &\cellcolor{gray!35} $4.4\times10^{-7}$  &\cellcolor{gray!15}  $3.5[4.9]\times10^{-6}$  & $7.9[11~]\times10^{-4}$  & $2.0[2.8]\times10^{-2}$ & - & - & 1500~(450)
\\\hline
$C_{s\nu1(2)}^{S,\alpha\beta}$  &\cellcolor{gray!35} $4.7\times10^{-6}$  & - &\cellcolor{gray!15}  $2.6[3.7]\times10^{-4}$ & $4.4[6.3]\times10^{-3}$  & - & - & 460~(52)
\\\hline
$C_{u\nu}^{T,\alpha\beta}$ &\cellcolor{gray!35}  $1.7\times10^{-6}$ & - & - & - &  $2.3\times10^{-2}$ & -& 780
\\\hline
$C_{d\nu}^{T,\alpha\beta}$ &\cellcolor{gray!35}  $0.9\times10^{-6}$  & - & - & - &$2.3\times10^{-2}$ & - & 1100
\\\hline
$C_{s\nu}^{T,\alpha\beta}$  &\cellcolor{gray!35}  $7.7\times10^{-6}$ & - & - & - & - & $1.0\times10^{-2}$ &360
\\\hline\hline 
$C_{NNF}^{\alpha\beta}$  & - & - & - & - & $1.4\times10^{-1}$ &\cellcolor{gray!35}  $7.8\times10^{-2}$ & 13
\\\hline
$C_{uN1(2)}^{S,\alpha\beta}$   & - &\cellcolor{gray!35}  $1.6[2.3]\times10^{-6}$ & $3.7[5.2]\times10^{-4}$  & $9.3[13~]\times10^{-3}$ & - & - &790[660]
\\\hline
$C_{dN1(2)}^{S,\alpha\beta}$   & - &\cellcolor{gray!35}  $3.5[4.9]\times10^{-6}$ & $7.9[11~]\times10^{-4}$ & $2.0[2.8]\times10^{-2}$ & - & - & 540[450]
\\\hline
$C_{sN1(2)}^{S,\alpha\beta}$   & - & - &\cellcolor{gray!35}  $2.6[3.7]\times10^{-4}$ & $4.4[6.3]\times10^{-3}$ & - &- & 62[52]
\\\hline
$C_{uN}^{T,\alpha\beta}$  & - & - & - & - &\cellcolor{gray!35}  $2.3\times10^{-2}$ & - & 6.6
\\\hline
$C_{dN}^{T,\alpha\beta}$   & - & - & - & - &\cellcolor{gray!35}  $2.3\times10^{-2}$ & - & 6.6
\\\hline
$C_{sN}^{T,\alpha\beta}$  & - & - & - & - & - &\cellcolor{gray!35}  $1.0\times10^{-2}$ & 10
\\\hline\hline 
$C_{u\nu N1(2)}^{V,\alpha\beta}$   &\cellcolor{gray!35}  $2.4\times10^{-6}$ & - &- & -  &\cellcolor{gray!15} $1.0\times10^{-1}$ & - & 650~(3.2)
\\\hline
$C_{d\nu N1(2)}^{V,\alpha\beta}$  &\cellcolor{gray!35}  $2.4\times10^{-6}$  & - &- & -  &\cellcolor{gray!15}  $1.0\times10^{-1}$ & - &650~(3.2)
\\\hline
$C_{s\nu N1(2)}^{V,\alpha\beta}$  & - & - &- & -  & - &\cellcolor{gray!35}  $4.4\times10^{-2}$ & 4.8
\\
\hline
\end{tabular}
}
\caption{Constraints on the Wilson coefficients of the LNV operators in the LNEFT. For the scalar type operators, the numbers outside [inside] the square bracket indicate the case with the neutrino flavors $\alpha\neq\beta [\alpha=\beta]$. Note that the Wilson coefficients $C_{q\nu1(2)}^{S,\tau\tau}$ in the first sector and $C_{q\nu N1(2)}^{V,\tau \beta}$ in the last sector can not be constrained by CE$\nu$NS. The strongest constraints on them are from the meson decays marked by the light gray cells, and the corresponding effective scale is shown in parentheses (\dots). }
\label{tab:LNEFTLNV}
\end{table}

Next, Tables~\ref{tab:LNEFTLNC} and \ref{tab:LNEFTLNV} show the constraints on the Wilson coefficients of LNEFT from the CE$\nu$NS process and meson invisible decays in the limit of massless RH neutrinos. The neutrino flavors $\alpha, \beta$ are arbitrary unless they are specified for $C^{V,\alpha\beta}_{q\nu 1(2)}$ or taken to be $\alpha=e, \mu$ in CE$\nu$NS process. In the LNV case, for the scalar-type operators with symmetric neutrino flavors, the numbers outside and inside the square bracket in Table~\ref{tab:LNEFTLNV} indicate the case with different neutrino flavors $\alpha\neq\beta$ and identical flavors $\alpha=\beta$, respectively. The gray cell displays the strongest constraint for each Wilson coefficient. One can see that the vector meson decays provide the sole bound on the particular flavor components $C_{u(d)\nu 1(2)}^{V,\tau\tau}$ and $C_{s\nu 1(2)}^{V,\alpha\beta}$ in the LNC case and $C_{s\nu N1(2)}^{V,\alpha\beta}$ in the LNV case. The coefficients without active neutrino degree of freedom, such as $C_{qN1(2)}^{V,\alpha\beta}$ in the LNC case and $C_{NNF}^{\alpha\beta}, C_{qN1(2)}^{S,\alpha\beta}, C_{qN}^{T,\alpha\beta}$ in the LNV case, can only be constrained by meson decays. The CE$\nu$NS process places the most stringent bound on all remaining Wilson coefficients with $\alpha=e,\mu$. The remaining WCs with $\alpha=\tau$ can not be constrained by CE$\nu$NS and we highlight the strongest constraints by meson decays in light gray. In the last columns of Tables~\ref{tab:LNEFTLNC} and \ref{tab:LNEFTLNV}, we also show the effective scale derived from the strongest constraint for each Wilson coefficient. The effective scales shown in parentheses correspond to the WCs in light gray for $\alpha=\tau$.

We then include the one-loop QCD/QED running result for the LNEFT Wilson coefficients from the chiral symmetry breaking scale to the electroweak scale and match them to SMNEFT at the electroweak scale in order to constrain new physics using the matching conditions from Table~\ref{match}. In Table~\ref{tab:SMNEFT} we display the constraints on the Wilson coefficients $C_i(\Lambda_{\rm EW})$ associated with the relevant dim-6 and dim-7 SMNEFT operators from the strongest limits of the corresponding LNEFT WCs in the gray sectors of Tables~\ref{tab:LNEFTLNC} and \ref{tab:LNEFTLNV}. By further assuming $\Lambda_{\rm NP}\equiv |C_i(\Lambda_{\rm EW})|^{1/(4-d)}$ with $d$ being the SMNEFT operator dimension, the constraints on the Wilson coefficients are also converted into the limits on the NP scale in units of the SM Higgs vev. The most stringent bounds on the NP scale are
\begin{eqnarray}
  \Lambda_{\rm NP}^{\rm dim-6}&=&\big(C_{NB}^{\alpha\beta}+C_{NW}^{\alpha\beta}\big)^{-{1\over2}}> 41\ v \ (\alpha=e, \mu)\;, \\
  \Lambda_{\rm NP}^{\rm dim-7}&=&\big(2C_{LHB}^{\alpha\beta}+C_{LHW}^{\beta\alpha}-C_{LHW}^{\alpha\beta}\big)^{-{1\over3}}>11 \ v \ (\alpha,\beta=e,\mu,\tau)\;,
\end{eqnarray}
from the corresponding dipole operators in LNEFT and
\begin{eqnarray}
  \Lambda_{\rm NP}^{\rm dim-6}&=&\big(C_{QuNL}^{11\beta\alpha}\big)^{-1/2}>6.0 \ v \ (\alpha=e, \mu)\;, \\
  \Lambda_{\rm NP}^{\rm dim-6}&=&\big(C_{QuNL}^{11\beta\alpha}\big)^{-1/2}>2.9 \ v \ (\alpha=\tau)\;, \\
  \Lambda_{\rm NP}^{\rm dim-7}&=&\big(C_{\bar QuLLH}^{11\alpha\beta}+C_{\bar QuLLH}^{11\beta\alpha}\big)^{-1/3}>2.9 \ v \ (\alpha=e, \mu)\;, \\
  \Lambda_{\rm NP}^{\rm dim-7}&=&\big(C_{\bar QuLLH}^{11\alpha\beta}\big)^{-1/3}>2.0 \ v \ (\alpha=\beta=\tau)\;,
\end{eqnarray}
from neutrino-quark operators in LNEFT.
Note that in this paper we work in a mixed mass-flavor basis for neutrinos.
In a realistic Seesaw model, due to the $N-\nu$ mixing, the interactions of sterile neutrinos with SM
particles exhibit a small mixing and are further suppressed. This generally leads to a natural suppression of the Wilson coefficient and thus relaxes the constraints on new particles mediating the effective interactions.

\begin{table}
\resizebox{\linewidth}{!}{
\renewcommand{\arraystretch}{1.14}
\begin{tabular}{|c|c|c|c|c|c|}
\hline
\cellcolor{gray!35}dim-6 SMNEFT WC & $[v^{-2}]$ & $\Lambda_{\rm NP}\equiv\left|C_i\right|^{-{1\over2}}[v]$  &\cellcolor{gray!35} dim-6 SMNEFT WC & $[v^{-2}]$ & $\Lambda_{\rm NP}\equiv\left|C_i\right|^{-{1\over2}}[v]$
\\\hline
$C_{lq}^{(1),ee11}+C_{lq}^{(3),ee11},C_{lu}^{ee11}$ &0.90 & 1.1 &
$C_{lq}^{(1),ee11}-C_{lq}^{(3),ee11},C_{ld}^{ee11}$& 0.82 & 1.1
\\\hline
$C_{lq}^{(1),e\mu11}+C_{lq}^{(3),e\mu11},C_{lu}^{e\mu11}$ & 0.26 & 2.0 &
$C_{lq}^{(1),e\mu11}-C_{lq}^{(3),e\mu11},C_{ld}^{e\mu11}$ & 0.22 & 2.1
\\\hline
 $C_{lq}^{(1),e\tau11}+C_{lq}^{(3),e\tau11},C_{lu}^{e\tau11}$ &0.36 & 1.7 &
 $C_{lq}^{(1),e\tau11}-C_{lq}^{(3),e\tau11},C_{ld}^{e\tau11}$ & 0.34 & 1.7
\\\hline
 $C_{lq}^{(1),\mu\mu11}+C_{lq}^{(3),\mu\mu11},C_{lu}^{\mu\mu11}$& 0.90 & 1.1 &
 $C_{lq}^{(1),\mu\mu11}-C_{lq}^{(3),\mu\mu11},C_{ld}^{\mu\mu11}$ & 0.82 & 1.1
\\\hline
 $C_{lq}^{(1),\mu\tau11}+C_{lq}^{(3),\mu\tau11},C_{lu}^{\mu\tau11}$ & 0.32  & 1.8 &
 $C_{lq}^{(1),\mu\tau11}-C_{lq}^{(3),\mu\tau11},C_{ld}^{\mu\tau11}$  & 0.30 & 1.8
\\\hline
$C_{NB}^{\alpha\beta}+C_{NW}^{\alpha\beta}, \alpha=e,\mu$ & $6.0\times 10^{-4}$ & 41  &
& &
\\\hline
 $C_{LNQd}^{\alpha\beta11}-{1\over2}C_{LdQN}^{\alpha11\beta},\alpha=e,\mu$ & $3.2\times 10^{-2}$ & 5.6 &
 $C_{LNQd}^{\tau\beta11}-{1\over2}C_{LdQN}^{\tau11\beta}$ & 0.25  & 2.0
\\\hline
$C_{LNQd}^{\alpha\beta22}-{1\over2}C_{LdQN}^{\alpha22\beta},\alpha=e,\mu$ & 0.34 & 1.7  &
$C_{LNQd}^{\tau\beta22}-{1\over2}C_{LdQN}^{\tau22\beta}$ & 19& 0.23
\\\hline
$C_{QuNL}^{11\beta\alpha},\alpha=e,\mu$  &$2.8\times10^{-2}$ &6.0 &
$C_{QuNL}^{11\beta\tau}$  & 0.12& 2.9
\\\hline
$C_{LdQN}^{\alpha11\beta},\alpha=e,\mu$& 0.13 & 2.8 &
$C_{LdQN}^{\alpha22\beta},\alpha=e,\mu$    &1.1& 1.0
\\
\hline\hline
\cellcolor{gray!35} dim-7 SMNEFT WC & $[v^{-3}]$ & $\Lambda_{\rm NP}\equiv\left|C_i\right|^{-{1\over3}}[v]$  &\cellcolor{gray!35} dim-7 SMNEFT WC & $[v^{-3}]$ & $\Lambda_{\rm NP}\equiv\left|C_i\right|^{-{1\over3}}[v]$
\\\hline
$2C_{LHB}^{\alpha\beta}+C_{LHW}^{\beta \alpha}-C_{LHW}^{\alpha\beta}$ & $8.5\times 10^{-4}$ & 11&
& &
\\\hline
 $C_{\bar dLQLH1}^{1\alpha1\beta}+C_{\bar dLQLH1}^{1\beta1\alpha},\alpha=e,\mu $  &0.09 & 2.2 &
$C_{\bar dLQLH1}^{1\tau1\tau}$  & 0.51 & 1.3
\\\hline
 $C_{\bar dLQLH1}^{2\alpha2\beta}+C_{\bar dLQLH1}^{2\beta2\alpha},\alpha=e,\mu $ & 0.97 & 1.0&
$C_{\bar dLQLH1}^{2\tau2\tau}$ & 38 & 0.3
 \\\hline
$C_{\bar QuLLH}^{11\alpha\beta}+C_{\bar QuLLH}^{11\beta\alpha},\alpha=e,\mu $  & 0.04 & 2.9 &
$C_{\bar QuLLH}^{11\tau\tau}$ & 0.12 & 2.0
\\\hline
$C_{\bar dLQLH1}^{1\alpha1\beta}-C_{\bar dLQLH1}^{1\beta1\alpha} $ & 1.5&  0.88 &
$C_{\bar dLQLH1}^{2\alpha2\beta}-C_{\bar dLQLH1}^{2\beta2\alpha} $ & 13 & 0.43
\\\hline
$C_{QNuH}^{1\alpha\beta1}+C_{QNuH}^{1\beta\alpha1}$ & 0.33[0.48] & $1.5[1.3]$ &
$C_{QNdH}^{1\alpha\beta1}+C_{QNdH}^{1\beta\alpha1}$ & 0.72[1.0]& $1.1[1.0]$
\\\hline
$C_{uQNH}^{11\alpha\beta}$ & 0.08[0.12] & 2.3 [2.0]&
$C_{dQNH}^{11\alpha\beta}$  & 0.18[0.25] & 1.8[1.6]
\\\hline 
$C_{QNLH1}^{11\beta\alpha},\alpha=e,\mu$  & 0.21 & 1.7 &
$C_{QNLH1}^{11\beta\alpha}-C_{QNLH2}^{11\beta\alpha},\alpha=e,\mu$ & 0.21 & 1.7
\\\hline
$C_{uNLH}^{11\beta\alpha},\alpha=e,\mu$  & 0.21 & 1.7 &
$C_{dNLH}^{11\beta\alpha},\alpha=e,\mu$ & 0.21 & 1.7
\\
\hline
\end{tabular}
}
\caption{Constraints on the Wilson coefficients of the relevant dim-6 and dim-7 SMNEFT from the strongest limits for the corresponding LNEFT WCs in the gray sector of Table~\ref{tab:LNEFTLNC} and Table~\ref{tab:LNEFTLNV}, where $v\simeq 246$ GeV is SM Higgs vacuum expectation value. For the dim-7 scalar type operators, the numbers outside [inside] the square bracket indicate the case with the neutrino flavors $\alpha\neq\beta [\alpha=\beta]$.
}
\label{tab:SMNEFT}
\end{table}

Besides the CE$\nu$NS and meson invisible decays, the search for dark matter in events with an energetic jet and large missing energy at the Large Hadron Collider (LHC) can also place constraints on the quark-neutrino interactions.
The interpretations of the LHC search for dark matter assumed simplified models with a mediator between the SM and the dark sector. The relevant limits can be applied for the constraints on the energy scale in the neutrino-quark interactions by taking nearly massless sterile neutrinos.
Assuming the neutrino-quark interactions mediated by a colored scalar, the mono-jet search excluded the mediator mass up to 1.67 TeV~\cite{Aaboud:2017phn} or 1.4 TeV~\cite{Sirunyan:2017jix}. A vector mediator is excluded up to a mass of 3.1 TeV~\cite{Aaboud:2017phn} assuming the couplings to be unity.

Although this work is focused on the quark-neutrino neutral current interactions, we find some of the involved SMNEFT operators,
which are obtained by matching using the results in Table~\ref{match},
also contribute to the quark-lepton charged interactions. There could be complementary constraints on them from low energy nuclear-level processes like beta decay and neutrinoless double beta ($0\nu\beta\beta$) decay. Specifically, the Wilson coefficients $C_{QuNL}^{11\beta e}, C_{LNQd}^{e\beta 11}, C_{LdQN}^{e11\beta }$ contribute to the $\beta$ decay directly, while $C_{\bar QuLLH}^{11\alpha\beta}, C_{\bar dLQLH1(2)}^{1\alpha1\beta}$ to the $0\nu\beta\beta$ decay via the long distance contribution which is mediated by neutrinos.
From Ref.~\cite{Bischer:2019ttk}, the constraints from beta decay translate into a limit on the NP scale in our basis
\begin{eqnarray}
\left(C_{QuNL}^{11e e}\right)^{-{1\over2}} \gtrsim 9{\rm ~TeV}\;,~
\left(C_{LNQd}^{ee11}-C_{LdQN}^{e11e}/2\right)^{-{1\over2}} \gtrsim 9{\rm ~TeV}\;,~
\left(C_{LdQN}^{e11e }\right)^{-{1\over2}} \gtrsim 400{\rm ~GeV}\;.
\end{eqnarray}
Similarly,
using the results in Ref.~\cite{Cirigliano:2017djv,Liao:2019tep,Bolton:2019wta}, $0\nu\beta\beta$ decay constrains the NP scale for the latter LNV operators as
\begin{eqnarray}
\left(C_{\bar QuLLH}^{11ee}\right)^{-{1\over3}}\ \gtrsim 100{\rm ~TeV}\;,~
\left(C_{\bar dLQLH1}^{1e1e}\right)^{-{1\over3}} \gtrsim 100{\rm ~TeV}\;,~
\left(C_{\bar dLQLH2}^{1e1e}\right)^{-{1\over3}} \gtrsim 100{\rm ~TeV}\;.~
\end{eqnarray}
For massive sterile neutrinos $N$ the LNC  operators  ${\cal O}_{LNQd}^{e\beta 11}, {\cal O}_{LdQN}^{e11\beta }$ can also contribute to the $0\nu\beta\beta$ decay via the mass mechanism. The authors of Ref.~\cite{Dekens:2020ttz} performed a detailed calculation of the contribution of sterile neutrinos to the $0\nu\beta\beta$ process in an effective field theory framework.
In particular the scale of the operator $\calO_{LdQN}^{e11\beta}$ is constrained to be $(C_{LdQN}^{e11\beta})^{-1/2}\gtrsim10$ TeV for sterile neutrino masses $m_N\in [0.1~\mathrm{MeV},100~\mathrm{GeV}]$.
However, for $m_N<1~\mathrm{keV}$ the CE$\nu$NS process places a stronger constraint on this operator.
Nevertheless, compared with Table~\ref{tab:SMNEFT}, one can see the charged current processes indeed give more stringent bounds on the NP scale associated with the relevant operators. We leave a detailed study of the charged current processes to future work~\cite{CC}.

\section{Conclusions}
\label{sec:Con}

We investigate the complementarity of the CE$\nu$NS process and meson invisible decay in constraining generic neutrino interactions with RH neutrinos in effective field theories. The interactions between quarks and left-handed SM neutrinos and/or right-handed neutrinos are first described by the LNEFT between the electroweak scale and the chiral symmetry breaking scale. We complete the independent operator basis for the LNEFT up to dim-6 by including both the LNC and LNV operators. We translate the bounds on the LNEFT Wilson coefficients from the COHERENT observation and calculate the branching fractions of light meson invisible decays. Finally, we include the one-loop QCD/QED running for the LNEFT Wilson coefficients from chiral symmetry breaking scale to the electroweak scale. The bounds on the LNEFT Wilson coefficients are then matched up to the SMNEFT to constrain new physics above the electroweak scale.
We summarize our main conclusions in the following
\begin{itemize}
\item In the LNC case, the vector meson invisible decays provide the sole but weak constraint on $C^{V,\tau\tau}_{q\nu 1(2), {\rm NP}}$, $C^{V,\alpha\beta}_{s\nu 1(2)}$ and $C^{V,\alpha\beta}_{qN1(2)}$. The LNEFT cutoff scale is $2-4$ GeV. CE$\nu$NS places the most stringent bound on the other vector LNEFT operators as well as $C_{\nu NF}^{\alpha\beta}$, $C^{S,\alpha\beta}_{q\nu N1(2)}$ and $C^{T,\alpha\beta}_{q\nu N}$ with $\alpha=e, \mu$. The WCs $C_{\nu NF}^{\tau\beta}$, $C^{S,\tau\beta}_{q\nu N1(2)}$ and $C^{T,\tau\beta}_{q\nu N}$ can only be constrained by meson decay.
\item In the LNV case, the meson invisible decays provide the sole constraint on $C_{NNF}^{\alpha\beta}$, $C^{S,\alpha\beta}_{qN1(2)}$, $C^{T,\alpha\beta}_{qN}$ and $C^{V,\alpha\beta}_{s\nu N1(2)}$. CE$\nu$NS gives the most stringent constraint on $C_{\nu\nu F}^{\alpha\beta}$, $C^{T,\alpha\beta}_{q\nu}$ and the components with $\alpha=e,\mu$ in $C^{S,\alpha\beta}_{q\nu 1(2)}$ and $C^{V,\alpha\beta}_{u(d)\nu N1(2)}$. The WCs $C^{S,\tau\beta}_{q\nu 1(2)}$ and $C^{V,\tau\beta}_{u(d)\nu N1(2)}$ can only be constrained by meson decay.
\item The most stringent bounds on the NP scale in SMNEFT are
\begin{eqnarray}
  \Lambda_{\rm NP}^{\rm dim-6}&=&\big(C_{NB}^{\alpha\beta}+C_{NW}^{\alpha\beta}\big)^{-{1\over2}}>41 \ v \simeq 10 \ {\rm TeV} \ (\alpha=e,\mu)\;, \nonumber \\
  \Lambda_{\rm NP}^{\rm dim-7}&=&\big(2C_{LHB}^{\alpha\beta}+C_{LHW}^{\beta\alpha}-C_{LHW}^{\alpha\beta}\big)^{-{1\over3}}>11\ v \simeq 2.7 \ {\rm TeV} \ (\alpha,\beta=e,\mu,\tau)\;, \nonumber
\end{eqnarray}
from the corresponding dipole operators in LNEFT and
\begin{eqnarray}
  \Lambda_{\rm NP}^{\rm dim-6}&=&\big(C_{QuNL}^{11\beta\alpha}\big)^{-1/2}>6.0 \ v \simeq 1.5 \ {\rm TeV} \ (\alpha=e, \mu)\;,  \nonumber\\
  \Lambda_{\rm NP}^{\rm dim-6}&=&\big(C_{QuNL}^{11\beta\alpha}\big)^{-1/2}>2.9 \ v \simeq 0.7 \ {\rm TeV} \ (\alpha=\tau)\;, \nonumber \\
  \Lambda_{\rm NP}^{\rm dim-7}&=&\big(C_{\bar QuLLH}^{11\alpha\beta}+C_{\bar QuLLH}^{11\beta\alpha}\big)^{-1/3}>2.9 \ v \simeq 0.7 \ {\rm TeV} \ (\alpha=e, \mu)\;,  \nonumber\\
  \Lambda_{\rm NP}^{\rm dim-7}&=&\big(C_{\bar QuLLH}^{11\alpha\beta}\big)^{-1/3}>2.0 \ v \simeq 0.5 \ {\rm TeV} \ (\alpha=\beta=\tau)\;, \nonumber
\end{eqnarray}
from neutrino-quark operators in LNEFT.
\end{itemize}
Finally, we comment on the UV-completions of the above EFT operators.
While the small mixing between active and sterile neutrinos leads to suppressed Wilson coefficients of the effective operators with neutrinos and quarks
in the conventional Seesaw models, Wilson coefficients are generally unsuppressed in models with additional interactions beyond the neutrino Yukawa coupling.

Simple extensions are UV models with an additional neutral $Z^\prime$ gauge bosons. This includes models of gauged U$(1)$ lepton number symmetry, where right-handed neutrinos are naturally present to cancel anomalies and the new gauge interaction will introduce new interactions of sterile neutrinos with other SM fermions which are not suppressed by the active-sterile mixing. Similarly, left-right symmetric theories introduce right-handed neutrinos with new gauge interactions.
See e.g.~Refs.~\cite{Wise:2014oea,Han:2019zkz,Miranda:2020zji} for recent studies of neutrino interactions with charged SM fermions within models with new $Z^\prime$ gauge bosons.
Also several classes of radiative neutrino mass models features large neutrino-quark interactions~\cite{Babu:2019mfe}.

Another possibility to produce effective quark-neutrino interactions from
a generation of $SU(2)_L$ doublet and singlet vector-like leptons
together with a singly-charged scalar has been discussed in Ref.~\cite{Chala:2020vqp}.
The simplest model that gives quark-neutrino interactions is a leptoquark which couples to a quark and a sterile neutrino (See e.g.~Ref.~\cite{Dorsner:2016wpm} for a recent review of leptoquarks.). This includes the SU$(2)_L$ singlet scalar leptoquarks $S_1$, $\bar S_1$ and vector leptoquarks $U_1$, $\bar U_1$ as well as the $SU(2)_L$ doublet scalar and vector leptoquarks $\tilde R_2$~\cite{Dekens:2020ttz} and $\tilde V_2$.
Finally, the minimal supersymmetric SM (MSSM) provides several new contributions to neutrino-quark operators. Within the MSSM with conserved R-parity, the effective operators can arise at the one-loop level~\cite{Bellazzini:2010gn}.

\section*{ACKNOWLEDGMENTS}
TL would like to thank Yi Liao and Cen Zhang for very useful discussion and communication.
TL is supported by the National Natural Science Foundation of China (Grant No. 11975129) and ``the Fundamental Research Funds for the Central Universities'', Nankai University (Grants No. 63196013, 63191522). XDM is supported by the MOST (Grant No. MOST 106-2112-M-002-003-MY3). MS acknowledges support by the Australian Research Council via the Discovery Project DP200101470.

\appendix

\section{The complete operator basis involving RH neutrinos $N$ in the LNEFT}
\label{sec:LNEFTbasis}

In this section we construct the complete and independent operator basis for the LNEFT involving RH neutrinos $N$ up to dim-6.
We work in the chiral basis and collectively denote the left- and right-handed down-type quarks as $d_{L}$ and $d_{R}$, the up-type quarks as $u_{L}$ and $u_{R}$, charged leptons as $e_{L}$ and $e_{R}$, and the SM left-handed neutrino fields as $\nu$ and the RH neutrinos as $N$, respectively. We drop the flavor indices for all of these fields for simplicity. For a fermion field $\psi$, its charge conjugation is defined via $\psi^C=C\bar\psi^{\T}$ where the matrix $C$ satisfies the relations $C^{\T}=C^\dagger=-C$ and $C^2=-1$. Except the up-type quarks with the total flavors $n_u=2$, the remaining charged fermions have $n_f=3$ flavors. We consider an arbitrary number $n_f$ of $N$ flavors.

At dim-5, it is easy to figure out that there are two independent non-hermitian operators
\begin{align}
\calO_{NNF}=&~(\overline{N^C}\sigma_{\mu\nu}N)F^{\mu\nu}, &
\calO_{\nu NF}=&~(\overline{\nu}\sigma_{\mu\nu}N)F^{\mu\nu}\;.
\end{align}
The full list of independent LNEFT operators with at least one RH neutrino $N$ at dim-6 is listed in Tables~\ref{tab:LNEFTA} and \ref{tab:LNEFTB}, where in the third and sixth columns in each table we also show the independent number of operators with flavors being considered. All those operators are classified in terms of the net number of the SM global baryon and lepton quantum numbers. An independent subset of lepton and baryon number conserving operators in LNEFT is given in Ref.~\cite{Chala:2020vqp}.

\begin{table}
\center
\resizebox{\linewidth}{!}{\footnotesize
\renewcommand{\arraystretch}{0.9}
\begin{tabular}{|l | l |  l|l | l | l|}
\hline
 Operator & Specific form & $\# (n_f,~n_u)$ &  Operator & Specific form & $\# (n_f,~n_u)$
\\
\hline
\hline
 \multicolumn{6}{|c|}{$(\Delta L, \Delta B)=(0,~0)$}
\\
\hline
\multicolumn{3}{|c|}{\cellcolor{gray!35}$(\overline{L}L)(\overline{R}R)$}
&
\multicolumn{3}{|c|}{\cellcolor{gray!35}$(\overline{R}R)(\overline{R}R)$}
\\
\hline
$\calO_{eN1}^V(\star\star)(\text{H})$ & $(\overline{e_L}\gamma_\mu e_L)(\overline{N}\gamma_\mu N)$ & $n_f^4 $ &
$\calO_{eN2}^V(\star\star)(\text{H})$ & $(\overline{e_R}\gamma_\mu e_R)(\overline{N}\gamma_\mu N)$ & $n_f^4$
\\
 $\calO_{dN1}^V(\star\star)(\text{H})$ & $(\overline{d_L}\gamma_\mu d_L)(\overline{N}\gamma_\mu N)$ & $n_f^4  $ &
 $\calO_{dN2}^V(\star\star)(\text{H})$ & $(\overline{d_R}\gamma_\mu d_R)(\overline{N}\gamma_\mu N)$ & $n_f^4 $
\\
$\calO_{uN1}^V(\star\star)(\text{H})$ & $(\overline{u_L}\gamma_\mu u_L)(\overline{N}\gamma_\mu N)$ & $n_f^2n_u^2  $ &
$\calO_{uN2}^V(\star\star)(\text{H})$ & $(\overline{u_R}\gamma_\mu u_R)(\overline{N}\gamma_\mu N)$ & $n_f^2n_u^2  $
\\
$\calO_{udeN1}^V(\star)$ & $(\overline{u_L}\gamma^\mu d_L)(\overline{e_R}\gamma_\mu N)$ & $n_f^3n_u $ &
$\calO_{udeN2}^V(\star)$ & $(\overline{u_R}\gamma^\mu d_R)(\overline{e_R}\gamma_\mu N)$ & $n_f^3n_u $
\\
$\calO_{\nu N}^V(\star\star)(\text{H})$ & $(\overline{\nu}\gamma_\mu \nu)(\overline{N}\gamma_\mu N)$ &
$ n_f^4 $ &
$\calO_{N}^V(\star\star\star\star)(\text{H})$ & $(\overline{N}\gamma_\mu N)(\overline{N}\gamma_\mu N)$ &
$ \frac{1}{4}n_f^2(n_f+1)^2 $
\\
\hline
\multicolumn{3}{|c|}{\cellcolor{gray!35}$(\overline{L}R)(\overline{L}R)$}
&
\multicolumn{3}{|c|}{\cellcolor{gray!35}$(\overline{R}L)(\overline{L}R)$}
\\
\hline
$\calO_{e\nu N1}^S(\star)$ & $(\overline{e_L}e_R)(\overline{\nu}N)$ & $n_f^4 $ &
$\calO_{e\nu N2}^S(\star)$ & $(\overline{e_R}e_L)(\overline{\nu}N)$ & $n_f^4 $
\\
$\calO_{e\nu N}^T(\star)$ & $(\overline{e_L}\sigma^{\mu\nu}e_R)(\overline{\nu}\sigma_{\mu\nu}N)$ & $n_f^4 $ &
& &
\\
$\calO_{d\nu N1}^S(\star)$ & $(\overline{d_L}d_R)(\overline{\nu}N)$ & $n_f^4 $ &
$\calO_{d\nu N2}^S(\star)$ & $(\overline{d_R}d_L)(\overline{\nu}N)$ & $n_f^4 $
\\
$\calO_{d\nu N}^T(\star)$ & $(\overline{d_L}\sigma^{\mu\nu}d_R)(\overline{\nu}\sigma_{\mu\nu}N)$ & $n_f^4 $ &  & &
\\
$\calO_{u\nu N1}^S(\star)$ & $(\overline{u_L}u_R)(\overline{\nu}N)$ & $n_f^2n_u^2$ &
$\calO_{u\nu N2}^S(\star)$ & $(\overline{u_R}u_L)(\overline{\nu}N)$ & $n_f^2n_u^2$
\\
 $\calO_{u\nu N}^T(\star)$ & $(\overline{u_L}\sigma^{\mu\nu}u_R)(\overline{\nu}\sigma_{\mu\nu}N)$ & $n_f^2n_u^2  $ &  & &
\\
$\calO_{udeN1}^S(\star)$ & $(\overline{u_L}d_R)(\overline{e_L}N)$ & $n_f^3n_u$ &
$\calO_{udeN2}^S(\star)$ & $(\overline{u_R}d_L)(\overline{e_L}N)$ & $n_f^3n_u$
\\
$\calO_{udeN}^T(\star)$ & $(\overline{u_L}\sigma^{\mu\nu}d_R)(\overline{e_L}\sigma_{\mu\nu}N)$ & $n_f^3n_u $ &  & &
\\
$\calO_{\nu N\nu N}^S(\star\star)$ & $(\overline{\nu} N)(\overline{\nu}N)$ & $\frac{1}{2} n_f^2(n_f^2+1)$ &
& &
\\
\hline
\hline
 \multicolumn{6}{|c|}{$(\Delta L, \Delta B)=(2,~0)$}
\\
\hline
\multicolumn{3}{|c|}{\cellcolor{gray!35}$(\overline{L}L)(\overline{R}R)$}
&
\multicolumn{3}{|c|}{\cellcolor{gray!35}$(\overline{R}R)(\overline{R}R)$}
\\
\hline
$\calO_{e\nu N1}^V(\star)$ & $(\overline{e_L}\gamma_\mu e_L)(\overline{\nu^C}\gamma_\mu N)$ & $n_f^4 $ &
$\calO_{e\nu N2}^V(\star)$ & $(\overline{e_R}\gamma_\mu e_R)(\overline{\nu^C}\gamma_\mu N)$ & $n_f^4  $
\\
$\calO_{d\nu N1}^V(\star)$ & $(\overline{d_L}\gamma_\mu d_L)(\overline{\nu^C}\gamma_\mu N)$ & $n_f^4$ &
$\calO_{d\nu N2}^V(\star)$ & $(\overline{d_R}\gamma_\mu d_R)(\overline{\nu^C}\gamma_\mu N)$ & $n_f^4$
\\
$\calO_{u\nu N1}^V(\star)$ & $(\overline{u_L}\gamma_\mu u_L)(\overline{\nu^C}\gamma_\mu N)$ & $n_f^2n_u^2$ &
$\calO_{u\nu N2}^V(\star)$ & $(\overline{u_R}\gamma_\mu u_R)(\overline{\nu^C}\gamma_\mu N)$ & $n_f^2n_u^2$
\\
$\calO_{dueN1}^V(\star)$ & $(\overline{d_L}\gamma_\mu u_L)(\overline{e_L^C}\gamma_\mu N)$ & $n_f^3n_u$ &
$\calO_{dueN2}^V(\star)$ & $(\overline{d_R}\gamma_\mu u_R)(\overline{e_L^C}\gamma_\mu N)$ & $n_f^3n_u$
\\
$\calO_{\nu\nu N}^V(\star)$ & $(\overline{\nu}\gamma_\mu \nu)(\overline{\nu^C}\gamma_\mu N)$ &
$ \frac{1}{2}n_f^3(n_f+1)$ &
$\calO_{N\nu N}^V(\star\star\star)$ & $(\overline{N}\gamma_\mu N)(\overline{\nu^C}\gamma_\mu N)$ &
$ \frac{1}{2}n_f^3(n_f+1)$
\\
\hline
\end{tabular}
}
\caption{Dim-6 operator basis involving RH neutrinos $N$ in LNEFT. Here all operators are non-hermitian expect those with a (H) in the first sector. The number of $\star$ after each operator indicates the number of the RH neutrinos involved in the same operator.}
\label{tab:LNEFTA}
\end{table}

\begin{table}[h!]
\centering
\resizebox{\linewidth}{!}{
\footnotesize
\renewcommand{\arraystretch}{0.9}
\begin{tabular}{|l | l |  l|l | l | l|}
\hline
 Operator & Specific form & $\# (n_f,~n_u)$ &  Operator & Specific form & $\# (n_f,~n_u)$
\\
\hline
\multicolumn{3}{|c|}{\cellcolor{gray!35}$(\overline{L}R)(\overline{L}R)$}
&
\multicolumn{3}{|c|}{\cellcolor{gray!35}$(\overline{R}L)(\overline{L}R)$}
\\
\hline
$\calO_{eN1}^S(\star\star)$ & $(\overline{e_L}e_R)(\overline{N^C}N)$ & $\frac{1}{2}n_f^3(n_f+1)$ &
$\calO_{eN2}^S(\star\star)$ & $(\overline{e_R}e_L)(\overline{N^C}N)$ & $\frac{1}{2}n_f^3(n_f+1)$
\\
$\calO_{eN}^T(\star\star)$ & $(\overline{e_L}\sigma_{\mu\nu}e_R)(\overline{N^C}\sigma^{\mu\nu}N)$ & $\frac{1}{2}n_f^3(n_f-1)$ &  & &
\\
\hline
$\calO_{dN1}^S(\star\star)$ & $(\overline{d_L}d_R)(\overline{N^C}N)$ & $\frac{1}{2}n_f^3(n_f+1)$ &
$\calO_{dN2}^S(\star\star)$ & $(\overline{d_R}d_L)(\overline{N^C}N)$ & $\frac{1}{2}n_f^3(n_f+1)$
\\
$\calO_{dN}^T(\star\star)$ & $(\overline{d_L}\sigma_{\mu\nu}d_R)(\overline{N^C}\sigma^{\mu\nu}N)$ & $\frac{1}{2}n_f^3(n_f-1)$ &  & &
\\
$\calO_{uN1}^S(\star\star)$ & $(\overline{u_L}u_R)(\overline{N^C}N)$ & $\frac{1}{2}n_u^2n_f(n_f+1)$  &
$\calO_{uN2}^S(\star\star)$ & $(\overline{u_R}u_L)(\overline{N^C}N)$ & $\frac{1}{2}n_f(n_f+1)n_u^2$
\\
$\calO_{uN}^T(\star\star)$ & $(\overline{u_L}\sigma_{\mu\nu}u_R)(\overline{N^C}\sigma^{\mu\nu}N)$ & $\frac{1}{2}n_u^2n_f(n_f-1)$ &  & &
\\
$\calO_{dueN1}^S(\star)$& $(\overline{d_L}u_R)(\overline{e_R^C}N)$ & $n_f^3n_u$ &
$\calO_{dueN2}^S(\star)$ & $(\overline{d_R}u_L)(\overline{e_R^C}N)$ & $n_f^3n_u$
\\
\cline{4-6}
$\calO_{dueN}^T(\star)$& $(\overline{d_L}\sigma^{\mu\nu} u_R)(\overline{e_R^C}\sigma_{\mu\nu} N)$ & $n_f^3n_u$ &
\multicolumn{3}{|c|}{\cellcolor{gray!35}$(\overline{R}L)(\overline{R}L)$}
\\
\cline{4-6}
$\calO_{\nu NN}^S(\star\star\star)$& $(\overline{\nu}N)(\overline{N^C}N)$ & $\frac{1}{3}n_f^2(n_f^2-1)$ &
$\calO_{N\nu \nu}^S(\star)$ & $(\overline{N}\nu )(\overline{\nu^C}\nu)$ & $\frac{1}{3}n_f^2(n_f^2-1)$
\\
\hline
\hline
 \multicolumn{6}{|c|}{$(\Delta L, \Delta B)=(4,~0)$}
\\
\hline
\multicolumn{3}{|c|}{\cellcolor{gray!35}$(\overline{L}R)(\overline{L}R)$}
&
\multicolumn{3}{|c|}{\cellcolor{gray!35}$(\overline{R}L)(\overline{L}R)$}
\\
\hline
$\calO_{N}^S(\star\star\star\star)$ &$(\overline{N^C}N)(\overline{N^C}N)$  & $\frac{1}{12}n_f^2(n_f^2-1)$ &
$\calO_{\nu N}^S(\star\star)$ & $(\overline{\nu^C}\nu)(\overline{N^C}N)$ &  $\frac{1}{4}n_f^2(n_f+1)^2$
\\
\hline
 \multicolumn{6}{|c|}{$(\Delta L, \Delta B)=(1,~-1)$}
\\
\hline
\multicolumn{3}{|c|}{\cellcolor{gray!35}$(\overline{R}R)(\overline{R}R)$}
&
\multicolumn{3}{|c|}{\cellcolor{gray!35}$(\overline{L}R)(\overline{L}R)$}
\\
\hline
$\calO_{dduN1}^V(\star)$ & $(\overline{d_R}\gamma^\mu d_L^C)(\overline{u_R}\gamma_\mu N)$ & $n_f^3n_u$ &
$\calO_{uddN1}^S(\star)$ & $(\overline{u_L}d_L^C)(\overline{d_L} N)$ & $n_f^3n_u$
\\
\cline{1-3}
\multicolumn{3}{|c|}{\cellcolor{gray!35}$(\overline{R}L)(\overline{L}R)$} & & &
\\
\cline{1-3}
$\calO_{dduN1}^S(\star)$ & $(\overline{d_R}d_R^C)(\overline{u_L} N)$ &$\frac{1}{2}n_f^2(n_f-1)n_u$ & & &
\\
\hline
\hline
\multicolumn{6}{|c|}{$(\Delta L, \Delta B)=(1,~1)$}
\\
\hline
\multicolumn{3}{|c|}{\cellcolor{gray!35}$(\overline{L}L)(\overline{R}R)$}
&
\multicolumn{3}{|c|}{\cellcolor{gray!35}$(\overline{L}R)(\overline{L}R)$}
\\
\hline
$\calO_{dduN2}^V(\star)$ & $(\overline{d_R^C}\gamma^\mu d_L)(\overline{u_L^C}\gamma_\mu N)$ & $n_f^3n_u$ &
$\calO_{uddN2}^V(\star)$ & $(\overline{u_R^C}d_R)(\overline{d_R^C} N)$ & $n_f^3n_u$
\\
\cline{1-3}
\multicolumn{3}{|c|}{\cellcolor{gray!35}$(\overline{R}L)(\overline{L}R)$}  & & &
\\
\cline{1-3}
$\calO_{dduN2}^S(\star)$ & $(\overline{d_L^C} d_L)(\overline{u_R^C}N)$ & $\frac{1}{2}n_f^2(n_f-1)n_u$ & & &
\\
\hline
\hline
\multicolumn{6}{|c|}{Total $\#=2331|_{B=0}^{L=0} + 2304|_{B=0}^{L=2} +84|_{B=0}^{L=4}+252|_{B=-1}^{L=1} +252|_{B=1}^{L=2}=5223$,~~ $(n_f,~n_u)=(3,~2)$}
\\
\hline
\end{tabular}
}
\caption{Continuation of Tab.~\ref{tab:LNEFTA}. }
\label{tab:LNEFTB}
\end{table}

\section{The SMNEFT operator basis at dim-6 and dim-7}
\label{sec:SMNEFTbasis}
Besides the SMEFT operators at dim-6~\cite{Grzadkowski:2010es} and dim-7~\cite{Lehman:2014jma,Liao:2016hru}, the SMNEFT also includes additional operators involving RH SM singlet fermions $N$. These operators with RH neutrino $N$ are classified in Ref.~\cite{Liao:2016qyd} and repeated in Table~\ref{tab:SMNEFT6} at dim-6 and Table~\ref{tab:SMNEFT7} at dim-7. For the dim-7 operators, by using the Fierz transformations here, we have rearranged some of the four-fermion operators given in Ref.~\cite{Liao:2016qyd} to have clear flavor symmetry and quark-lepton current structure. In addition, for the operators involving gauge field strength tensors, we accompany a corresponding gauge coupling constant for each involved field strength tensor.  Besides the operator basis involving RH neutrinos $N$ in Table~\ref{tab:SMNEFT6} and Table~\ref{tab:SMNEFT7}, in our matching calculation we also need the following relevant SMEFT dim-6 operators
\begin{align}
{\cal O}_{lq}^{(1)}=&(\overline{L}\gamma_\mu L)(\overline{Q}\gamma^\mu Q)\;, &
{\cal O}_{lq}^{(3)}=&(\overline{L}\gamma_\mu \tau^I L)(\overline{Q}\gamma^\mu \tau^IQ)\;, \nonumber
\\
{\cal O}_{lu}=&(\overline{L}\gamma_\mu L)(\overline{u}\gamma^\mu u)\;, &
{\cal O}_{ld}=&(\overline{L}\gamma_\mu L)(\overline{d}\gamma^\mu d)\;,
\end{align}
and also dim-7 operators
\begin{align}
{\cal O}_{LHB}=&g_1\epsilon_{i j} \epsilon_{m n}(\overline{L^{Ci}}\sigma_{\mu \nu} L^{m})H^{j} H^{n} B^{\mu \nu}\;, &
{\cal O}_{\bar dLQLH1}=&\epsilon_{i j} \epsilon_{m n}(\bar{d} L^{i})(\overline{Q^{Cj}} L^{m}) H^{n}\;,  \nonumber
\\
{\cal O}_{LHW}=&g_2\epsilon_{i j}(\epsilon\tau^I)_{m n}(\overline{L^{Ci}} \sigma_{\mu \nu} L^{m}) H^{j} H^{n} W^{I\mu \nu}\;, &
{\cal O}_{\bar QuLLH}=&\epsilon_{i j}(\bar{Q} u)(\overline{L^C} L^{i}) H^{j}\;,
\end{align}
where $g_{1,2}$ are the gauge coupling constants for the gauge groups $U(1)_Y$ and $SU(2)_L$, respectively.

\begin{table}
\center
\parbox{1\linewidth}{
\scriptsize
\centering
\setlength{\tabcolsep}{9.5pt}
\renewcommand{\arraystretch}{1.1}
\begin{tabular}{|c|c|c|c|c|c|}
\hline
\multicolumn{2}{|c|}{$\psi^2H^3(+\mbox{h.c.})$}
&\multicolumn{2}{|c|}{$(\overline{L}R)(\overline{L}R)(+\mbox{h.c.})$}
&\multicolumn{2}{|c|}{$(\overline{L}L)(\overline{R}R)$}
\\\hline
$\calO_{LNH}$ & $(\overline{L}N)\tilde{H}(H^\dagger H)$
& $\calO_{LNLe}$ & $(\overline{L}N)\epsilon(\overline{L}e)$
&$\calO_{LN}$ & $(\overline{L}\gamma^\mu L)(\overline{N}\gamma_\mu N)$
\\\cline{1-4}
\multicolumn{2}{|c|}{$\psi^2H^2D (+\mbox{h.c.})$}
&$\calO_{LNQd}$ & $(\overline{L}N)\epsilon(\overline{Q}d)$
&$\calO_{QN}$ & $(\overline{Q}\gamma^\mu Q)(\overline{N}\gamma_\mu N)$
\\\cline{1-2}\cline{5-6}
$\calO_{HN}(\text{H})$ &  $(\overline{N}\gamma^\mu N)(H^\dagger i \overleftrightarrow{D_\mu} H)$
& $\calO_{LdQN}$ & $(\overline{L}d)\epsilon(\overline{Q}N)$
&\multicolumn{2}{|c|}{\cellcolor{gray!35}$(\Delta L, \Delta B)=(4,~0)$}
\\\cline{3-6}
$\calO_{HNe}$ & $(\overline{N}\gamma^\mu e)({\tilde{H}}^\dagger i D_\mu H)$
& \multicolumn{2}{|c|}{$(\overline{R}R)(\overline{R}R)$}
&$\cellcolor{gray!35}\calO_{NNNN}$ &$\cellcolor{gray!35}(\overline{N^C}N)(\overline{N^C}N)$
\\\hline
\multicolumn{2}{|c|}{$\psi^2HX(+\mbox{h.c.})$}
&$\calO_{NN}$ & $(\overline{N}\gamma^\mu N)(\overline{N}\gamma_\mu N)$
&\multicolumn{2}{|c|}{$\cellcolor{gray!35}(\Delta L, \Delta B)=(1,~1)$}
\\\cline{1-2}\cline{5-6}
$\calO_{NB}$ & $g_1(\overline{L}\sigma_{\mu\nu}N)\tilde{H}B^{\mu\nu}$
&$\calO_{eN}$ & $(\overline{e}\gamma^\mu e)(\overline{N}\gamma_\mu N)$
&$\cellcolor{gray!35}\calO_{QQdN}$ &$\cellcolor{gray!35}\epsilon_{ij}\epsilon_{\alpha\beta\sigma}(\overline{Q^{i,C}_{\alpha}}Q^j_{\beta})(\overline{d_{\sigma}^C}N)$
\\
$\calO_{NW}$ &$g_2(\overline{L}\sigma_{\mu\nu}N)\tau^I\tilde{H}W^{I\mu\nu}$
&$\calO_{uN}$& $(\overline{u}\gamma^\mu u)(\overline{N}\gamma_\mu N)$
&$\cellcolor{gray!35}\calO_{uddN}$ &$\cellcolor{gray!35}\epsilon_{\alpha\beta\sigma}(\overline{u_{\alpha}^C}d_{\beta})(\overline{d_{\sigma}^C}N)$
\\\cline{1-2}\cline{5-6}
\multicolumn{2}{|c|}{$(\overline{L}R)(\overline{R}L)(+\mbox{h.c.})$}
&$\calO_{dN}$& $(\overline{d}\gamma^\mu d)(\overline{N}\gamma_\mu N)$
& &
\\\cline{1-2}
$\calO_{QuNL}$& $(\overline{Q}u)(\overline{N}L)$
&$\calO_{duNe}(+\mbox{h.c.})$& $ (\overline{d}\gamma^\mu u)(\overline{N}\gamma_\mu e)$
& &
\\
\hline
\end{tabular}
\caption{The basis of dim-6 operators involving RH neutrino $N$ in SMNEFT~\cite{Liao:2016hru}, where $\alpha,~\beta,~\sigma$ and $i,~j$ are $SU(3)_C$ and $SU(2)_L$ indices, respectively.}
\label{tab:SMNEFT6}
}
\vspace{0.5cm}
\\
\parbox{1\linewidth}{
\scriptsize
\centering
\setlength{\tabcolsep}{3pt} 
\renewcommand{\arraystretch}{1.1}
\begin{tabular}{|c|c|c|c|c|c|}
\hline
 \multicolumn{2}{|c|}{$N\psi H^3D$} &  \multicolumn{2}{|c|}{$N\psi^3D$}   &   \multicolumn{2}{|c|}{$N^2\psi^2H$}
\\\hline
$\mathcal{O}_{NL1}$ & $\epsilon_{ij}(\overline{N^C}\gamma_\mu L^i) (iD^\mu H^j)(H^\dagger H)$
&$\mathcal{O}_{eNLLD}$ & $\epsilon_{ij}(\overline{e}\gamma_\mu N)(\overline{L^{i,C}}i\overleftrightarrow{D}^\mu L^j)$
& $\mathcal{O}_{LNeH}$ & $(\overline{L}N)(\overline{N^C}e)H$
\\
$\mathcal{O}_{NL2}$&$\epsilon_{ij}(\overline{N^C}\gamma_\mu L^i)  H^j(H^\dagger i\overleftrightarrow{D^\mu} H)$
&  $\mathcal{O}_{duNeD}$& $(\overline{d}\gamma_\mu u)(\overline{N^C}i\overleftrightarrow{D}^\mu e)$
& $\mathcal{O}_{eLNH}$ & $H^\dagger(\overline{e}L)(\overline{N^C}N)$
\\\cline{1-2}
 \multicolumn{2}{|c|}{$N\psi H^2D^2$}   &  $\mathcal{O}_{QuNLD}$ & $(\overline{Q}i\overleftrightarrow{D}_\mu u)(\overline{N^C}\gamma^\mu L)$ &
$\mathcal{O}_{QNdH}$ & $(\overline{Q}N)(\overline{N^C}d)H$
\\\cline{1-2}
$\mathcal{O}_{NeD}$ &  $\epsilon_{ij}(\overline{N^C}\overleftrightarrow{D}^\mu e)(H^iD^\mu H^j)$
&  $\mathcal{O}_{dQNLD}$  & $\epsilon_{ij}(\overline{d}i\overleftrightarrow{D}_\mu Q^i )(\overline{N^C}\gamma^\mu L^j) $
&  $\mathcal{O}_{dQNH}$ & $H^\dagger(\overline{d}Q)(\overline{N^C}N)$
\\\cline{1-4}
\multicolumn{2}{|c|}{$N\psi H^2X$}
& \multicolumn{2}{|c|}{$N^2\psi^2D$}
& $\mathcal{O}_{QNuH}$ & $(\overline{Q}N)(\overline{N^C}u)\tilde{H}$
\\\cline{1-4}
$\mathcal{O}_{NeW}$ & $g_2(\epsilon\tau^I)_{ij}(\overline{N^C}\sigma^{\mu\nu}e)(H^iH^j)W^I_{\mu\nu}$
& $\mathcal{O}_{LND}$ & $(\overline{L}\gamma_\mu L)(\overline{N^C}i\overleftrightarrow{\partial}^\mu N)$
& $\mathcal{O}_{uQNH}$ & $\tilde{H}^\dagger(\overline{u}Q)(\overline{N^C}N)$
\\\cline{1-2}\cline{5-6}
\multicolumn{2}{|c|}{$N\psi HDX$}
&  $\mathcal{O}_{QND}$ & $(\overline{Q}\gamma_\mu Q)(\overline{N^C}i\overleftrightarrow{\partial}^\mu N)$
&  \multicolumn{2}{|c|}{$N^3\psi H$}
\\\cline{1-2}\cline{5-6}
$\mathcal{O}_{NLB1}$ & $g_1\epsilon_{ij}(\overline{N^C}\gamma^\mu L^i)(D^\nu H^j)B_{\mu\nu}$
& $\mathcal{O}_{eND}$ & $(\overline{e}\gamma_\mu e)(\overline{N^C}i\overleftrightarrow{\partial}^\mu N)$
& $\mathcal{O}_{LNNH}$  & $(\overline{L}N)(\overline{N^C}N)\tilde{H}$
\\
$\mathcal{O}_{NLB2}$ & $g_1\epsilon_{ij}(\overline{N^C}\gamma^\mu L^i)(D^\nu H^j)\tilde{B}_{\mu\nu}$
& $\mathcal{O}_{uND}$ & $(\overline{u}\gamma_\mu u)(\overline{N^C}i\overleftrightarrow{\partial}^\mu N)$
& $\mathcal{O}_{NLNH}$ & $\tilde{H}^\dagger(\overline{N}L)(\overline{N^C}N)$
\\\cline{5-6}
$\mathcal{O}_{NLW1}$ & $g_2(\epsilon \tau^I)_{ij}(\overline{N^C}\gamma^\mu L^i)(D^\nu H^j)W^I_{\mu\nu}$
& $\mathcal{O}_{dND}$ & $(\overline{d}\gamma_\mu d)(\overline{N^C}i\overleftrightarrow{\partial}^\mu N)$
&  \multicolumn{2}{|c|}{\cellcolor{gray!35}$\slashed{B}:~N\psi^3D~\&~N\psi^3H$}
\\\cline{3-6}
$\mathcal{O}_{NLW2}$ & $g_2(\epsilon \tau^I)_{ij}(\overline{N^C}\gamma^\mu L^i)(D^\nu H^j)\tilde{W}^I_{\mu\nu}$
&\multicolumn{2}{|c|}{$N^4D$}
& $\cellcolor{gray!35} \mathcal{O}_{uNdD}$ &\cellcolor{gray!35}$\epsilon_{\alpha\beta\sigma}(\overline{u}_{\alpha}\gamma_\mu N)(\overline{d}_{\beta}i\overleftrightarrow{D}^\mu d^C_{\sigma})$
\\\cline{1-4}
\multicolumn{2}{|c|}{$N^2H^4$}
& $\mathcal{O}_{NND}$ & $(\overline{N}\gamma_\mu N)(\overline{N^C}i\overleftrightarrow{\partial}^\mu N) $
& $\cellcolor{gray!35} \mathcal{O}_{dNQD}$ &\cellcolor{gray!35}$\epsilon_{ij}\epsilon_{\alpha\beta\sigma}(\overline{d}_{\alpha}\gamma_\mu N) (\overline{Q}_{i\beta}i\overleftrightarrow{D}^\mu Q_{j\sigma}^C)$
\\\cline{1-4}
$\mathcal{O}_{NH}$ & $(\overline{N^C}N)(H^\dagger H)^2$
&\multicolumn{2}{|c|}{$N\psi^3 H$}
&\cellcolor{gray!35}$\mathcal{O}_{QNdH}$ &\cellcolor{gray!35}$\epsilon_{ij}\epsilon_{\alpha\beta\sigma}(\overline{Q}_{i\alpha}N)(\overline{d}_\beta d^C_\sigma)\tilde{H}^j$
\\\cline{1-4}
\multicolumn{2}{|c|}{$N^2H^2D^2$}
& $\mathcal{O}_{LNLH}$ & $\epsilon_{ij}(\overline{L}\gamma_\mu L)(\overline{N^C}\gamma^\mu L^i)H^j$
 &\cellcolor{gray!35}$\mathcal{O}_{QNQH}$ &\cellcolor{gray!35}$\epsilon_{ij}\epsilon_{\alpha\beta\sigma}(\overline{Q}_{i\alpha}N)(\overline{Q}_{j\beta }Q^C_\sigma)H$
\\\cline{1-2}
$\mathcal{O}_{NHD1}$ & $(\overline{N^C}\overleftrightarrow{\partial}_\mu N)(H^\dagger \overleftrightarrow{D^\mu} H)$
& $\mathcal{O}_{QNLH1}$ & $\epsilon_{ij}(\overline{Q}\gamma_\mu Q)(\overline{N^C}\gamma^\mu L^i)H^j$
&\cellcolor{gray!35}$\mathcal{O}_{QNudH}$ &\cellcolor{gray!35}$\epsilon_{\alpha\beta\sigma}(\overline{Q}_{\alpha}N)(\overline{u}_\beta d^C_\sigma)H$
\\\cline{5-6}
    $\mathcal{O}_{NHD2}$        & $(\overline{N^C} N)(D_\mu H)^\dagger D^\mu H$
& $\mathcal{O}_{QNLH2}$ & $\epsilon_{ij}(\overline{Q}\gamma_\mu Q^i)(\overline{N^C}\gamma^\mu L^j)H$
&\multicolumn{2}{|c|}{$N^2X^2$}
\\\cline{1-2}\cline{5-6}
\multicolumn{2}{|c|}{$N^2H^2X$}
&$\mathcal{O}_{eNLH}$ & $\epsilon_{ij}(\overline{e}\gamma_\mu e )(\overline{N^C} \gamma^\mu L^i)H^j$
& $\mathcal{O}_{NB1}$ & $\alpha_1(\overline{N^C}N)B_{\mu\nu}B^{\mu\nu}$
\\\cline{1-2}
$\mathcal{O}_{NHB}$ & $g_1(\overline{N^C}\sigma_{\mu\nu}N)(H^\dagger H)B^{\mu\nu}$
& $\mathcal{O}_{dNLH}$   & $\epsilon_{ij}(\overline{d}\gamma_\mu d)(\overline{N^C}\gamma^\mu L^i)H^j$
& $\mathcal{O}_{NB2}$        & $\alpha_1(\overline{N^C}N)B_{\mu\nu}\tilde{B}^{\mu\nu}$
\\
$\mathcal{O}_{NHW}$       & $g_2(\overline{N^C}\sigma_{\mu\nu}N)(H^\dagger\tau^I H)W^{I\mu\nu}$
& $\mathcal{O}_{uNLH}$ & $\epsilon_{ij}(\overline{u}\gamma_\mu u)(\overline{N^C} \gamma^\mu L^i)H^j$
& $\mathcal{O}_{NW1}$    & $\alpha_2(\overline{N^C}N)W^I_{\mu\nu}W^{I\mu\nu}$
\\
&
&  $\mathcal{O}_{duNLH}$ & $\epsilon_{ij}(\overline{d}\gamma_\mu u)(\overline{N^C} \gamma^\mu L^i)\tilde{H}^j$
&  $\mathcal{O}_{NW2}$ & $\alpha_2(\overline{N^C}N)W^I_{\mu\nu}\tilde{W}^{I\mu\nu}$
\\
&
& $\mathcal{O}_{dQNeH}$ & $\epsilon_{ij}(\overline{d}Q^i)(\overline{N^C}e)H^j$
& $\mathcal{O}_{NG1}$      & $\alpha_3(\overline{N^C}N)G^A_{\mu\nu}G^{A\mu\nu}$
\\
&
& $\mathcal{O}_{QuNeH1}$ & $(\overline{Q}u)(\overline{N^C}e)H$
& $\mathcal{O}_{NG2}$      & $\alpha_3(\overline{N^C}N)G^A_{\mu\nu}\tilde{G}^{A\mu\nu}$
\\
&
& $\mathcal{O}_{QuNeH2}$ & $(\overline{Q}\sigma_{\mu\nu}u)(\overline{N^C}\sigma^{\mu\nu}e)H$
& &
\\\hline
\end{tabular}
\caption{The basis of dim-7 operators involving RH neutrino $N$ in SMNEFT, where all of the operators are non-hermitian with the net global quantum number $|\Delta L-\Delta B|=2$. Here $g_{1,2,3}$ are the gauge coupling constants for the gauge groups $U(1)_Y, SU(2)_L,SU(3)_C$, respectively, and $\alpha_i=g_i^2/(4\pi)$. }
\label{tab:SMNEFT7}
}
\end{table}

\section{Relations to other operator bases}
In this appendix we briefly summarize how our operator basis relates to other bases used in papers which we refer to in the main part of the text.

\subsection{Non-Standard Interactions}
\label{sec:NSI}
A commonly used operator basis are non-standard interactions (NSIs)~\cite{Wolfenstein:1977ue,Ohlsson:2012kf,Farzan:2017xzy} (Recent progress on NSI can be seen in Ref.~\cite{Dev:2019anc} and the references therein.), which describe the interactions of active neutrinos at low energies.

In particular, neutral-current interactions with quarks are described by\footnote{One possible underlying UV completion of these NSIs are models with a new neutral vector boson $Z^\prime$. See e.g.~Ref.~\cite{Flores:2020lji} for a recent study of COHERENT in the context of a $Z^\prime$ model.}
\begin{eqnarray}
\mathcal{L}_{\rm NSI}= -\sqrt{2}G_F \varepsilon_{\alpha\beta}^{qV}(\bar{\nu}_\alpha \gamma^\mu P_L \nu_\beta )\bar{q}\gamma_\mu q \; ,
\label{eq:NSI}
\end{eqnarray}
with $\varepsilon_{\alpha\beta}^{qV}=\varepsilon_{\beta\alpha}^{qV\ast}$.
The $\varepsilon$ parameterization is related to the NP contribution to the vector Wilson coefficients in our operator basis via
  \begin{equation}
    -\sqrt{2} G_F \epsilon_{\alpha\beta}^{qV} = \frac12\left( C_{q\nu1,{\rm NP}}^{V,\alpha\beta} + C_{q\nu2,{\rm NP}}^{V,\alpha\beta}\right)\;,
  \end{equation}
and to the $\xi_V$ parameter in Eq.~(\ref{eq:diffeq}) via
\begin{eqnarray}
\xi_V^2&= 4[(g_V^p+2\varepsilon_{\alpha\alpha}^{uV}+\varepsilon_{\alpha\alpha}^{dV})\mathbb{Z}F_p(Q^2)+(g_V^n+\varepsilon_{\alpha\alpha}^{uV}+2\varepsilon_{\alpha\alpha}^{dV})\mathbb{N}F_n(Q^2)]^2\nonumber \\
&+4\sum_{\beta\neq \alpha}|(2\varepsilon_{\beta\alpha}^{uV}+\varepsilon_{\beta\alpha}^{dV})\mathbb{Z}F_p(Q^2)+(\varepsilon_{\beta\alpha}^{uV}+2\varepsilon_{\beta\alpha}^{dV})\mathbb{N}F_n(Q^2)|^2\;,
\end{eqnarray}
with the SM couplings being
\begin{eqnarray}
g_V^p = {1\over 2}-2\sin^2\theta_W, \quad g_V^n = -{1\over 2} \;.\nonumber
\end{eqnarray}

\subsection{CD parameterization}
\label{sec:CD}
For the vector Wilson coefficients in LNV case, the relation to the quark-level $C_V^q$ parameter in Ref.~\cite{Chang:2020jwl} is
\begin{eqnarray}
C_V^{q\ast}-D_A^{q\ast}=-{1\over \sqrt{2}G_F}\Big(C_{q\nu N1}^{V,\alpha\beta}+C_{q\nu N2}^{V,\alpha\beta}\Big)\;.
\end{eqnarray}
For the scalar and tensor Wilson coefficients, we have the following relations to the quark-level parameters in Ref.~\cite{AristizabalSierra:2018eqm}
\begin{eqnarray}
C_S^q+iD_P^q={1\over \sqrt{2}G_F}\Big(C_{q\nu N1}^{S,\alpha\beta\ast}+C_{q\nu N2}^{S,\alpha\beta\ast}\Big)\;, \quad
C_T^q-iD_T^q={\sqrt{2}\over G_F}C_{q\nu N}^{T,\alpha\beta\ast}\;,
\end{eqnarray}
in the LNC case and
\begin{eqnarray}
C_S^q+iD_P^q={\sqrt{2}\over G_F}\Big(C_{q\nu 1}^{S,\alpha\beta}+C_{q\nu 2}^{S,\alpha\beta}\Big)\;, \quad C_T^q-iD_T^q=-{2\sqrt{2}\over G_F}C_{q\nu}^{T,\alpha\beta}\;,
\end{eqnarray}
in the LNV case.

\section{The matrix elements of neutrino scattering}
\label{sec:amplitude}

In the LNC case, the quark-level amplitudes for the scattering $\nu_\alpha(p_1) q(k_1)\to\nu_\beta/N_\beta(p_2) q(k_2)$ and $\bar\nu_\alpha(p_1) q(k_1)\to\bar\nu_\beta/\bar N_\beta(p_2) q(k_2)$ is given by
\begin{eqnarray}
\mathcal{M}(\nu_\alpha q\to \nu_\beta q)
&=&{1\over 2}(C_{q\nu 1}^{V,\beta\alpha}+C_{q\nu 2}^{V,\beta\alpha})(\overline{u_\nu}\gamma_\mu P_Lu_\nu)(\overline{q}\gamma^\mu q)
+\fbox{SD}\; , \nonumber \\
-\mathcal{M}(\bar\nu_\alpha q\to \bar \nu_\beta q)
&=&{1\over 2}(C_{q\nu1}^{V,\alpha\beta}+C_{q\nu2}^{V,\alpha\beta})(\overline{v_{\bar \nu}}P_R\gamma_\mu v_{\bar\nu})(\overline{q}\gamma^\mu q)+\fbox{SD}\; , \nonumber
\\
\mathcal{M}(\nu_\alpha q\to N_\beta q)
&=&{1\over 2}(C^{S,\alpha\beta*}_{q\nu N1}+C^{S,\alpha\beta*}_{q\nu N2})(\overline{u_N}P_Lu_\nu)(\overline{q}q)
+C^{T,\alpha\beta*}_{q\nu N}(\overline{u_N}\sigma_{\mu\nu}P_Lu_\nu)(\overline{q}\sigma^{\mu\nu}q)\nonumber \\
&+&i{2eQ_q\over q^2}C_{\nu NF}^{\alpha\beta*}(\overline{u_N}\sigma_{\mu\nu}P_Lu_\nu)(\overline{q}\gamma^\mu t^\nu q)+\fbox{SD}\;, \nonumber \\
-\mathcal{M}(\bar\nu_\alpha q\to \bar N_\beta q)
&=&{1\over 2}(C^{S,\alpha\beta}_{q\nu N1}+C^{S,\alpha\beta}_{q\nu N2})(\overline{v_{\bar \nu}}P_Rv_{\bar N})(\overline{q}q)
+C^{T,\alpha\beta}_{q\nu N}(\overline{v_{\bar \nu}}P_R\sigma_{\mu\nu}v_{\bar N})(\overline{q}\sigma^{\mu\nu}q)\nonumber \\
&+&i{2eQ_q\over q^2}C_{\nu NF}^{\alpha\beta}(\overline{v_{\bar \nu}}P_R\sigma_{\mu\nu}v_{\bar N})(\overline{q}\gamma^\mu t^\nu q)+\fbox{SD} \; ,
\end{eqnarray}
where \fbox{SD} stands for spin-dependent terms and the exchanged 4-momentum $q=p_1-p_2=k_2-k_1$.

In the LNV case, the quark-level amplitudes for the scattering $\nu_\alpha(p_1) q(k_1)\to\bar\nu_\beta/\bar N_\beta(p_2) q(k_2)$ and $\bar\nu_\alpha(p_1) q(k_1)\to\nu_\beta/N_\beta(p_2) q(k_2)$ are
\begin{eqnarray}
\mathcal{M}(\nu_\alpha q\to \bar\nu_\beta q)
&=&(C^{S,\alpha\beta}_{q\nu 1}+C^{S,\alpha\beta}_{q\nu 2})(\overline{v^C_{\bar \nu}}P_Lu_\nu)(\overline{q}q)
-2C^{T,\alpha\beta}_{q\nu}(\overline{v^C_{\bar \nu}}\sigma_{\mu\nu}P_Lu_\nu)(\overline{q}\sigma^{\mu\nu}q)\nonumber \\
&-& i{4eQ_q\over q^2}C_{\nu \nu F}^{\alpha\beta}(\overline{v_{\bar \nu}^C}\sigma_{\mu\nu}P_Lu_\nu)(\overline{q}\gamma^\mu t^\nu q)+\fbox{SD}\; , \nonumber\\
\mathcal{M}(\bar\nu_\alpha q\to \nu_\beta q)
&=&(C^{S,\alpha\beta\ast}_{q\nu 1}+C^{S,\alpha\beta\ast}_{q\nu 2})(\overline{v_{\bar \nu}}P_Ru^C_\nu)(\overline{q}q)
-2C^{T,\alpha\beta\ast}_{q\nu}(\overline{v_{\bar \nu}}P_R\sigma_{\mu\nu}u^C_\nu)(\overline{q}\sigma^{\mu\nu}q)\nonumber \\
&-& i{4eQ_q\over q^2}C_{\nu \nu F}^{\alpha\beta*}(\overline{v_{\bar \nu}}P_R\sigma_{\mu\nu}u_\nu^C)(\overline{q}\gamma^\mu t^\nu q)+\fbox{SD}\; ,\nonumber
\\
\mathcal{M}(\nu_\alpha q\to \bar N_\beta q)
&=&-{1\over 2}(C_{q\nu N1}^{V,\alpha\beta}+C_{q\nu N2}^{V,\alpha\beta})(\overline{v^C_{\bar N}}\gamma_\mu P_Lu_\nu)(\overline{q}\gamma^\mu q)
+\fbox{SD}\; ,\nonumber
\\
\mathcal{M}(\bar\nu_\alpha q\to N_\beta q)
&=&-{1\over 2}(C_{q\nu N1}^{V,\alpha\beta*}+C_{q\nu N2}^{V,\alpha\beta*})(\overline{v_{\bar \nu}}\gamma_\mu P_Lu^C_N)(\overline{q}\gamma^\mu q)
+\fbox{SD}\; .
\end{eqnarray}
The nucleon-level matrix elements for the scattering of an active antineutrino off a nucleus, $\bar\nu_\alpha(p_1) {\cal N}(k_1)\to\bar\nu_\beta/\bar N_\beta(p_2) {\cal N}(k_2)$, are given by
\begin{eqnarray}
-\mathcal{M}(\bar\nu_\alpha {\cal N}\to \bar\nu_\beta {\cal N})&=&{1\over 2}C_{{\cal N}\nu}^{V,\alpha\beta}(\overline{v_{\bar \nu}}P_R\gamma_\mu v_{\bar\nu})\bar {\cal N}\gamma^\mu{\cal N} \;, \nonumber
\\
-\mathcal{M}(\bar\nu_\alpha{\cal N}\to \bar N_\beta{\cal N})&=&{1\over 2}C^{S,\alpha\beta}_{{\cal N}\nu N}(\overline{v_{\bar \nu}}P_Rv_{\bar N})\bar {\cal N}{\cal N} +
C^{T,\alpha\beta}_{{\cal N}\nu N}(\overline{v_{\bar \nu}}P_R\sigma_{\mu\nu}v_{\bar N})\bar{\cal N}\sigma^{\mu\nu}{\cal N} \;,
\nonumber\\
-\mathcal{M}(\bar\nu_\alpha {\cal N}\to \bar N_\beta {\cal N})&=&i{eG_F\over q^2}A_{M\nu NF}^{\alpha\beta}(\overline{v_{\bar \nu}}P_R\sigma_{\mu\nu}v_{\bar N})\overline{{\cal N}}\gamma^\mu t^\nu {\cal N} \; ,
\end{eqnarray}
for LNC interactions and the corresponding matrix elements for LNV interactions are
\begin{eqnarray}
\mathcal{M}(\bar\nu_\alpha{\cal N}\to \nu_\beta{\cal N})&=&C^{S,\alpha\beta\ast}_{{\cal N}\nu}(\overline{v_{\bar{\nu}}}P_Ru^C_\nu)\bar {\cal N}{\cal N}
-2C^{T,\alpha\beta\ast}_{{\cal N}\nu}(\overline{v_{\bar{\nu}}}P_R\sigma_{\mu\nu}u^C_\nu)\bar{\cal N}\sigma^{\mu\nu}{\cal N} \;, \nonumber
\\
\mathcal{M}(\bar\nu_\alpha {\cal N}\to N_\beta {\cal N})&=&-{1\over 2}C_{{\cal N}\nu N}^{V,\alpha\beta\ast}(\overline{v_{\bar{\nu}}}\gamma_\mu P_Lu^C_N)\bar {\cal N}\gamma^\mu{\cal N} \;,
 \nonumber \\
\mathcal{M}(\bar\nu_\alpha {\cal N}\to \nu_\beta {\cal N})&=&
-i{eG_F\over q^2}A_{M\nu \nu F}^{\alpha\beta*}(\overline{v_{\bar \nu}}P_R\sigma_{\mu\nu}u_\nu^C)\overline{{\cal N}}\gamma^\mu t^\nu {\cal N}\; .
\end{eqnarray}

\section{The matrix elements of meson invisible decays}
\label{sec:invdecay}

For the quark-level process of $q\bar{q}\to {\rm inv}_1(k_1){\rm inv}_2(k_2)$, the LNC amplitudes are
\begin{eqnarray}
 \mathcal{M}(q\bar{q}\to \nu_\alpha\bar{\nu}_\beta)&=&  \left(C_{q\nu 1}^{V,\alpha\beta} \overline{q}\gamma_\mu P_L q + C_{q\nu 2}^{V,\alpha\beta} \overline{q}\gamma_\mu P_R q\right) \overline{u_\nu} \gamma^\mu P_L v_{\bar \nu} \;,
\nonumber\\
\mathcal{M}(q\bar{q}\to N_\alpha\bar{N}_\beta)&=& \left(C_{qN 1}^{V,\alpha\beta} \overline{q}\gamma_\mu P_L q + C_{qN 2}^{V,\alpha\beta} \overline{q}\gamma_\mu P_R q\right) \overline{u_N} \gamma^\mu P_R v_{\bar N} \;,
\nonumber\\
\mathcal{M}(q\bar{q}\to \nu_\alpha\bar{N}_\beta)&=& \left( C_{q\nu N1}^{S,\alpha\beta}\overline{q} P_R q  + C_{q\nu N2}^{S,\alpha\beta}\overline{q} P_L q \right)\overline{u_\nu} P_R v_{\bar N}\nonumber \\
&+&  \left( C_{q\nu N}^{T,\alpha\beta}\overline{q} \sigma_{\mu\nu}P_R q -i 2eQ_q C_{\nu NF}^{\alpha\beta}{(k_1+k_2)_\nu\over (k_1+k_2)^2} \overline{q}\gamma_\mu q\right) \overline{u_\nu}\sigma^{\mu\nu} P_R v_{\bar N} \;,
\nonumber\\
  \mathcal{M}(q\bar{q}\to \bar{\nu}_\alpha N_\beta)&=& \left(C_{q\nu N1}^{S,\alpha\beta\ast}\overline{q} P_L q + C_{q\nu N2}^{S,\alpha\beta\ast}\overline{q} P_R q \right) \overline{u_N} P_L v_{\bar \nu}\nonumber \\
&+& \left( C_{q\nu N}^{T,\alpha\beta\ast}\overline{q} \sigma_{\mu\nu}P_L q -i 2eQ_q C_{\nu NF}^{\alpha\beta\ast}{(k_1+k_2)_\nu\over (k_1+k_2)^2}\overline{q}\gamma_\mu q \right)\overline{u_N}\sigma^{\mu\nu} P_L v_{\bar \nu}\;,
\end{eqnarray}
The LNV amplitudes with $\Delta L=-2$ are
\begin{eqnarray}
\mathcal{M}(q\bar{q}\to \bar{\nu}_\alpha\bar{\nu}_\beta)&=& 2\left(C_{q\nu 1}^{S,\alpha\beta}\overline{q} P_L q +C_{q\nu 2}^{S,\alpha\beta}\overline{q} P_R q \right)\overline{v^C_{\bar \nu}} P_L v_{\bar \nu^\prime}
\nonumber \\
 &+&2\left(C_{q\nu}^{T,\alpha\beta}\overline{q} \sigma_{\mu\nu}P_L q -i 2eQ_qC_{\nu\nu F}^{\alpha\beta}{(k_1+k_2)_\nu\over (k_1+k_2)^2}\overline{q}\gamma_\mu q \right)\overline{v^C_{\bar\nu}}\sigma^{\mu\nu} P_L v_{\bar \nu^\prime}\;,
\nonumber \\
\mathcal{M}(q\bar{q}\to \bar{N}_\alpha\bar{N}_\beta)&=& 2\left(C_{qN 1}^{S,\alpha\beta}\overline{q} P_L q + C_{qN 2}^{S,\alpha\beta}\overline{q} P_R q\right) \overline{v^C_{\bar N}}  P_R v_{\bar N^\prime}
\nonumber \\
&+&2\left(C_{qN}^{T,\alpha\beta}\overline{q} \sigma_{\mu\nu}P_R q -i2 eQ_qC_{NN F}^{\alpha\beta}{(k_1+k_2)_\nu\over (k_1+k_2)^2}\overline{q}\gamma_\mu q\right) \overline{v^C_{\bar N}}\sigma^{\mu\nu}P_R v_{\bar N^\prime}\;,
\nonumber \\
\mathcal{M}(q\bar{q}\to \bar{\nu}_\alpha\bar{N}_\beta)&=& -\left(C_{q\nu N1}^{V,\alpha\beta}\overline{q}\gamma_\mu P_L q +C_{q\nu N2}^{V,\alpha\beta}\overline{q}\gamma_\mu P_R q\right) \overline{v^C_{\bar N}} \gamma^\mu P_Lv_{\bar \nu} \;,
\end{eqnarray}
where $v_{\bar \nu} (v_{\bar N})$ and $v_{\bar \nu^\prime} (v_{\bar N^\prime})$ are the spinors of anti-neutrinos $\bar \nu_\alpha (\bar N_\alpha)$ and $\bar \nu_\beta (\bar N_\beta)$, respectively. The amplitudes with $\Delta L=2$ are
\begin{eqnarray}
\mathcal{M}(q\bar{q}\to \nu_\alpha \nu_\beta)&=& 2\left(C_{q\nu 1}^{S,\alpha\beta\ast}\overline{q} P_R q +C_{q\nu 2}^{S,\alpha\beta\ast}\overline{q} P_L q\right)\overline{u_\nu} P_Ru^C_{\nu^\prime}
 \nonumber \\
&+&2\left(C_{q\nu}^{T,\alpha\beta\ast}\overline{q} \sigma_{\mu\nu}P_R q -i2eQ_qC_{\nu\nu F}^{\alpha\beta\ast}{(k_1+k_2)_\nu\over (k_1+k_2)^2}\overline{q}\gamma_\mu q \right) \overline{u_\nu} \sigma^{\mu\nu}P_R u^C_{\nu^\prime} \;,
 \nonumber \\
\mathcal{M}(q\bar{q}\to N_\alpha N_\beta)&=& 2\left(C_{qN 1}^{S,\alpha\beta\ast}\overline{q} P_R q +C_{qN 2}^{S,\alpha\beta\ast}\overline{q} P_L q\right)\overline{u_N} P_L u^C_{N^\prime}
  \nonumber\\
&+&2\left(C_{qN}^{T,\alpha\beta\ast}\overline{q} \sigma_{\mu\nu}P_L q -i2 eQ_qC_{NN F}^{\alpha\beta\ast}{(k_1+k_2)_\nu\over (k_1+k_2)^2}\overline{q}\gamma_\mu q \right)\overline{u_N} \sigma^{\mu\nu} P_Lu^C_{N^\prime} \;,
 \nonumber \\
 \mathcal{M}(q\bar{q}\to \nu_\alpha N_\beta)&=&-\left(C_{q\nu N1}^{V,\alpha\beta\ast}\overline{q}\gamma_\mu P_L q + C_{q\nu N2}^{V,\alpha\beta\ast}\overline{q}\gamma_\mu P_R q \right)\overline{u_\nu} \gamma^\mu P_Lu^C_{N} \;,
\end{eqnarray}
where $u_{\bar \nu} (u_{\bar N})$ and $u_{\bar \nu^\prime} (u_{\bar N^\prime})$ are the spinors of neutrinos $\nu_\alpha (N_\alpha)$ and $\nu_\beta (N_\beta)$, respectively.

We list the individual matrix elements for pseudoscalar invisible decays to neutrinos
\begin{align}
 -{\cal M}({P\to \nu_\alpha\bar\nu_\beta})=&{i\over 2}f_P^q\left( C_{q\nu 1}^{V,\alpha\beta}-C_{q\nu 2}^{V,\alpha\beta}\right)\overline{u_\nu} \slashed{p}P_L v_{\bar \nu}=0\;, \nonumber
\\
-{\cal M}({P\to N_\alpha \bar N_\beta})=&{i\over 2}f_P^q\left( C_{qN 1}^{V,\alpha\beta}-C_{qN 2}^{V,\alpha\beta}\right)\overline{u_N} \slashed{p} P_R v_{\bar N}\;, \nonumber
\\
-{\cal M}({P\to \nu_\alpha \bar N_\beta})=&i{h_P^q\over 4m_q} \left(C_{q\nu N1}^{S,\alpha\beta}-C_{q\nu N2}^{S,\alpha\beta}\right)\overline{u_\nu} P_R v_{\bar N}\;,\nonumber
\\
{\cal M}({P\to \bar \nu_\alpha N_\beta})=&i{h^q_P \over 4m_q} \left(C_{q\nu N1}^{S,\alpha\beta*}-C_{q\nu N2}^{S,\alpha\beta*}\right)\overline{u_N} P_L v_{\bar \nu}\;,\nonumber
\\
\mathcal{M}({P\to\bar{\nu}_\alpha\bar{\nu}_\beta})=&i{h^q_P \over 2m_q}\left( C_{q\nu 1}^{S,\alpha\beta}-C_{q\nu 2}^{S,\alpha\beta}\right)\overline{v^C_{\bar \nu}} P_L v_{\bar \nu^\prime}\;, \nonumber
\\
-\mathcal{M}({P\to{\nu}_\alpha{\nu}_\beta})=&i{h^q_P \over 2m_q}\left( C_{q\nu 1}^{S,\alpha\beta*}-C_{q\nu 2}^{S,\alpha\beta*}\right)\overline{u_{\nu}}P_R u^C_{\nu^\prime}\;, \nonumber
\\
\mathcal{M}({P\to \bar{N}_\alpha\bar{N}_\beta})=&i{h^q_P \over 2m_q}\left( C_{qN 1}^{S,\alpha\beta}-C_{qN 2}^{S,\alpha\beta}\right)\overline{v^C_{\bar N}} P_Rv_{\bar N^\prime}\;, \nonumber
\\
-\mathcal{M}({P\to{N}_\alpha {N}_\beta})=&i{h^q_P \over 2m_q}\left( C_{qN 1}^{S,\alpha\beta*}-C_{qN 2}^{S,\alpha\beta*}\right)\overline{u_{N}} P_L u^C_{N^\prime}\;, \nonumber
\\
\mathcal{M}(P\to \bar{\nu}_\alpha\bar{N}_\beta)=&if_P^q\left(C_{q\nu N1}^{V,\alpha\beta}-C_{q\nu N2}^{V,\alpha\beta}\right)\overline{v^C_{\bar N}} \slashed{p}P_L v_{\bar \nu} \;, \nonumber
\\
\mathcal{M}(P\to \nu_\alpha N_\beta)=&if_P^q\left(C_{q\nu N1}^{V,\alpha\beta\ast}-C_{q\nu N2}^{V,\alpha\beta\ast}\right)\overline{u_\nu} \slashed{p} P_L u^C_{N} \;,
\end{align}
where in each amplitude the quark label $q$ is summed over the first three light quarks $(u,d,s)$ implicitly
and $u_{\bar \nu} (u_{\bar N})$ and $u_{\bar \nu^\prime} (u_{\bar N^\prime})$ are the spinors of neutrinos $\nu_\alpha (N_\alpha)$ and $\nu_\beta (N_\beta)$, respectively.
By evaluating the squared matrix elements, we find the following results for the vector and scalar currents
\begin{eqnarray}
|\overline{u_1}\slashed{p} P_{L/R} u_2|^2 =m_P^2\left[m_1^2+m_2^2- {(m_1^2-m_2^2)^2\over m_P^2}\right],~
|\overline{u_1} P_{L/R} u_2|^2=m_P^2\left[1 -{m_1^2+m_2^2 \over m_P^2} \right].
\label{eq:relation1}
\end{eqnarray}
One can see that the above relation holds for any projection operator and it
is true for either particles or antiparticles in the final states. It also
applies for neutrino bilinears with charge-conjugate fields, since
$u(p,s)=C\bar v(p,s)^T$ and $v(p,s)=C\bar u(p,s)^T$. Taking all this together, we obtain the branching ratio in Eq.~\eqref{eq:BrPinv}.

Similarly the decay matrix elements of vector meson $V$ are given by
\begin{align}
\mathcal{M}(V\to \nu_\alpha\bar{\nu}_\beta)=&m_V{f_V^q\over 2}\left(C_{q\nu 1}^{V,\alpha\beta}+ C_{q\nu 2}^{V,\alpha\beta}\right) \overline{u_\nu}\gamma_\mu P_Lv_{\bar \nu}\epsilon_V^\mu\;, \nonumber\\
\mathcal{M}(V\to N_\alpha\bar{N}_\beta)=&m_V{f_V^q\over 2} \left(C_{qN 1}^{V,\alpha\beta}+ C_{qN 2}^{V,\alpha\beta}\right) \overline{u_N} \gamma_\mu P_R v_{\bar N}\epsilon_V^\mu \;, \nonumber\\
\mathcal{M}(V\to \nu_\alpha\bar{N}_\beta)=&i2\left(f_V^{T,q}C_{q\nu N}^{T,\alpha\beta} -eQ_q{f_V^q\over m_V}C_{\nu NF}^{\alpha\beta}\right) \overline{u_\nu}\sigma^{\mu\nu}P_R v_{\bar N}\epsilon_V^\mu p^\nu  \;, \nonumber\\
\mathcal{M}(V\to \bar{\nu}_\alpha N_\beta)=&i2\left(f_V^{T,q}C_{q\nu N}^{T,\alpha\beta\ast}-eQ_q{f_V^q\over m_V}C_{\nu NF}^{\alpha\beta\ast} \right)\overline{u_N}\sigma_{\mu\nu} P_Lv_{\bar \nu}\epsilon_V^\mu p^\nu\;,\nonumber\\
\mathcal{M}(V\to \bar{\nu}_\alpha\bar{\nu}_\beta)=&i4\left(f_V^{T,q}C_{q\nu}^{T,\alpha\beta}-eQ_q{f_V^q\over m_V}C_{\nu\nu F}^{\alpha\beta}\right)\overline{v^C_{\bar\nu}}\sigma_{\mu\nu}P_Lv_{\bar \nu^\prime}\epsilon_V^\mu p^\nu\;, \nonumber\\
\mathcal{M}(V\to \bar{N}_\alpha\bar{N}_\beta)=&i4\left(f_V^{T,q}C_{qN}^{T,\alpha\beta}-eQ_q{f_V^q\over m_V}C_{NN F}^{\alpha\beta}\right) \overline{v^C_{\bar N}}\sigma_{\mu\nu} P_Rv_{\bar N^\prime}\epsilon_V^\mu p^\nu\;,\nonumber\\
-\mathcal{M}(V\to \bar{\nu}_\alpha\bar{N}_\beta)=&m_V{f_V^q\over 2}\left(C_{q\nu N1}^{V,\alpha\beta} +C_{q\nu N2}^{V,\alpha\beta}\right) \overline{v^C_{\bar N}}\gamma_\mu P_Lv_{\bar \nu}\epsilon_V^\mu \;,\nonumber
\\
\mathcal{M}(V\to \nu_\alpha \nu_\beta)=&i4\left(f_V^{T,q}C_{q\nu}^{T,\alpha\beta\ast}-eQ_q{f_V^q\over m_V}C_{\nu\nu F}^{\alpha\beta\ast}\right) \overline{u_\nu} \sigma_{\mu\nu}P_Ru^C_{\nu^\prime}\epsilon_V^\mu p^\nu\;, \nonumber\\
\mathcal{M}(V\to N_\alpha N_\beta)=&i4\left(f_V^{T,q}C_{qN}^{T,\alpha\beta\ast}-eQ_q{f_V^q\over m_V}C_{NN F}^{\alpha\beta\ast} \right)\overline{u_N} \sigma_{\mu\nu}P_Lu^C_{N^\prime}\epsilon_V^\mu p^\nu\;, \nonumber\\
-\mathcal{M}(V\to \nu_\alpha N_\beta)=&m_V{f_V^q\over 2}\left(C_{q\nu N1}^{V,\alpha\beta\ast}+ C_{q\nu N2}^{V,\alpha\beta\ast}\right)\overline{u_\nu} \gamma_\mu P_Lu^C_{N}\epsilon_V^\mu \;,
\end{align}
where we again sum over quark flavor $q=u,d,s$ implicitly.
By evaluating the squared matrix elements, we find the following results for the vector and tensors currents
\begin{align}
\frac13\sum_{\rm pol}\left|\bar u_1\gamma_\mu P_{L/R} u_2\epsilon^\mu_V\right|^2
=&{2\over 3}m_V^2\left[1-{m_1^2+m_2^2\over 2m_V^2}-{(m_1^2-m_2^2)^2\over 2m_V^4}\right]\;,\nonumber
\\
\frac13\sum_{\rm pol}\left|\bar u_1\sigma_{\mu\nu} P_{L/R} u_2\epsilon^\mu_Vp^\nu\right|^2
=&{1\over 3}m^4_V\left[1+ {m_1^2+m_2^2\over m_V^2}-2{ (m_1^2-m_2^2)^2\over m_V^4}\right]\;,
\end{align}
after averaging over the initial polarizations of the vector meson $V$. Combining the above results, we obtain the branching ratio given in the main part of the text in Eq.~\eqref{eq:BrVinv}.

\bibliographystyle{JHEP}
\bibliography{refs}

\end{document}